%% file: JIneson.tex
\documentclass[a4paper,fleqn,usenatbib]{mnras} 

\usepackage{mathptmx}
\usepackage{txfonts}

\usepackage[T1]{fontenc}
\usepackage{ae,aecompl}

\usepackage{graphicx}	
\usepackage{mathrsfs}
\usepackage{times}

\title[The dynamics and energetics of FRII radio galaxies]{A representative survey of the dynamics and energetics of FRII radio galaxies}

\author[J. Ineson et al.]{
J. Ineson,$^{1}$\thanks{E-mail: J.Ineson@soton.ac.uk}
J. H. Croston,$^{1}$
M. J. Hardcastle,$^{2}$
and B. Mingo$^{3}$
\\
$^{1}$School of Physics and Astronomy, University of Southampton, Southampton S17 1BJ, UK\\
$^{2}$Centre for Astrophysics Research. School of Physics, Astronomy and Mathematics, University of Hertfordshire, Hatfield, Hertfordshire AL10 9AB, UK\\
$^{3}$Department of Physics and Astronomy, University of Leicester, University Road, Leicester LE1 7RH, UK
}

\date{Accepted XXX. Received YYY; in original form ZZZ}

\pubyear{2016}

\begin{document}
\label{firstpage}
\pagerange{\pageref{firstpage}--\pageref{lastpage}}
\maketitle

\begin{abstract} 
We report the first large, systematic study of the dynamics and energetics of a representative sample of FRII radio galaxies with well-characterized group/cluster environments. We used X-ray inverse-Compton and radio synchrotron measurements to determine the internal radio-lobe conditions, and these were compared with external pressures acting on the lobes, determined from measurements of the thermal X-ray emission of the group/cluster. Consistent with previous work, we found that FRII radio lobes are typically electron-dominated by a small factor relative to equipartition, and are over-pressured relative to the external medium in their outer parts. These results suggest that there is typically no energetically significant proton population in the lobes of FRII radio galaxies (unlike for FRIs), and so for this population, inverse-Compton modelling provides an accurate way of measuring total energy content and estimating jet power.  We estimated the distribution of Mach numbers for the population of expanding radio lobes, finding that at least half of the radio galaxies are currently driving strong shocks into their group/cluster environments. Finally, we determined a jet power--radio luminosity relation for FRII radio galaxies based on our estimates of lobe internal energy and Mach number. The slope and normalisation of this relation are consistent with theoretical expectations, given the departure from equipartition and environmental distribution for our sample.  \end{abstract}




\section{Introduction}\label{sec:intro}

The jets of radio-loud AGN consist of plasma drawn at relativistic speeds from the central regions of the AGN. The plasma is transported into the surrounding galaxy group or cluster where it forms into lobes, which displace the intra-cluster medium (ICM). Cavities carved in the ICM by the lobes have been observed in clusters out to redshifts greater than 0.5 \citep{hla12}, while detailed studies of local sources have shown complex substructures in the ICM (e.g. \citealt{kar02,fab03}) and shock fronts driven by the lobe expansion (e.g. \citealt{cro09,she11}). These features provide direct evidence that the radio jets are disturbing the ICM on a large scale and transferring energy from the AGN to the ICM. Estimates of the energy required to create the cavities around the lobes, and hence of the energy to be dissipated into the ICM, vary with the method used (e.g. \citealt{git12}), but enthalpies tend to lie between $10^{55}$ and $10^{61}$~erg depending on the richness of the cluster (e.g. \citealt{bir04}). This has been estimated to be sufficient to offset the ICM cooling and star formation that is predicted by evolutionary models but not seen in observations (e.g. \citealt{dun06,raf06}).

The dynamics of the lobes of radio-loud AGN are dependent on their pressure relative to that of the surrounding ICM \citep{scu74}. In Scheuer's models A and B the lobes are highly over-pressured and so expand at supersonic rates; in model C they are near pressure balance and so expand more gently and are capable of being moulded by the ICM. The latter model is more in keeping with observed lobe shapes, but the tip must remain over-pressured for the lobe to continue to grow and so there may still be supersonic movement out through the ICM.

If the expansion is supersonic, the lobes will input energy through shocks as well as through gas displacement (e.g. \citealt{bir04,hla12}). Lobe shocks have been seen in some observational studies, and estimates of their Mach numbers range from 1.2 to 8.4 (e.g. \citealt{wil06,bir08,cro09,she11a,cro11,wor12,kra12,nul13}). However, in many cases the uncertainties in the methods used to estimate Mach numbers mean that these are likely to be lower limits and so the true values could be higher. The rate of lobe expansion is therefore important in determing heat input into the ICM. \citet{san07} have made a detailed study of ripples propagating through the Perseus cluster, and suggest that 20--40~per~cent of the cavity power goes into sound waves which carry the energy outwards from the cavities and heat the ICM. For the more powerful FRII lobes, estimates of the energy input to the environment that are based on cavity volume are likely to underestimate the total energy input much more severely and so the usual $pV$ estimates of cavity power should be taken as lower limits (e.g. \citealt{git12,hak13,per14}). 

The relativistic leptons in the radio jets emit synchrotron radiation across a range of radio frequencies. The total energy emitted depends on both the particle and magnetic field energies, and we cannot separate these factors and determine the internal lobe conditions from just the radio flux observations (e.g. \citealt{lon11}, Section 16.5). We can however calculate the lobe conditions required to produce the observed flux by assuming that the two factors make similar contributions to the total energy -- the assumption of equipartition. This is similar to the minimum energy conditions required to produce the observed flux and so we can obtain a lower limit on the internal lobe pressure \citep{bur56}.

A second source of emission is inverse Compton (IC), where the relativistic particles boost photons in the lobes to X-ray frequencies and above, and this can be seen in X-ray observations of the lobes of many FRII sources. The dominant photon field is from CMB photons in the body of the lobes, with a small contribution from synchrotron self-Compton (SSC) (e.g. \citealt{har02b}). SSC is stronger at lobe hot-spots, in the knots of FRI jets and very occasionally in small, compact lobes where photon densities and electron energies are higher \citep{har04b,kat05,har10}. Some IC emission from nuclear photons has also been observed in a few lobes and when strong this could make an important contribution to the IC emission. 3C~207, for example, has much stronger X-ray emission near the nucleus than in the rest of the lobe \citep{bru02}. However, in other potential examples of nuclear emission the effect is not so clearly visible and any contribution to the overall X-ray flux in the lobe is thought to be small (e.g. \citealt{cro04,bel04,cro05b}).

If IC emission is observed across the lobe, this allows us to estimate the electron density and so gain a better estimate of internal lobe conditions than the assumption of equipartition. A number of observational studies and simulations have considered how the internal conditions of lobes differ from equipartition; their inferred particle content; how their internal pressure compares with that of the ICM; and how energy is transferred from the lobe to the ICM.

IC X-ray emission has not been detected from FRI lobes and so their lobe pressures are generally calculated assuming equipartition. These pressure estimates tend to be less than that of the surrounding ICM (e.g. \citealt{har98b,wor00,dun05}) which indicates the presence of a population of non-radiating particles to boost the pressure. \citet{cro08a} looked at the morphologies of a sample of FRI galaxies and found that those where the jets were in direct contact with the ICM showed a larger pressure deficit, suggesting that the additional particles are likely to come from entrainment of material from the ICM. Some detailed studies of local sources (e.g. \citealt{har10,cro14}) use X-ray observations to obtain upper limits on the lobe IC flux which places limits on the population of relativistic leptons. This population is insufficient to provide the required lobe pressures, suggesting there must also be a significant population of baryons. Measurements of external and internal conditions along the jet show that these are most likely to come from entrainment.

FRII sources tend to be more distant so there are fewer detailed studies, but they also appear under-pressured with respect to the environment when the pressure is estimated assuming equipartition (e.g. \citealt{har00}). Unlike FRI lobes, however, many FRII lobes show X-ray emission indicative of IC and when this used to better model the electron energy density in the lobes (assuming that the lobes are electron-dominated), the lobes tend to appear close to pressure balance with their environment (e.g. \citealt{har02b,cro04,kat05,cro05b,bel07}). This suggests that the assumption that the lobes are predominantly populated by relativistic electrons is likely to be reasonable; there may still be some lower energy baryons present but they can only account for a small amount of the energy budget (e.g. \citealt{har10}).

There are however two major assumptions made in the synchrotron and IC calculations that we cannot resolve from observations -- that the dominant relativistic particles are electrons/positrons, and the proportion of the lobe that is filled by the particles. As discussed by \citet{har00}; \citet{har02b} and \citet{cro05b}, if these assumptions are incorrect then, assuming that the synchrotron-emitting leptons are interspersed with non-radiating particles, the internal pressures of the lobes are likely to be under-estimated. This would result in lobes that are more highly over-pressured with respect to their environment. 

As described above, lobe-tip shocks have been observed in some studies, but the temperature variations across the shock can only be isolated for strong, nearby sources. Mach numbers can however be estimated from the ratio between internal and external lobe pressure (e.g. \citealt{wor06}). This method does however neglect the ram pressure, so Mach numbers obtained in this way should be regarded as lower limits, but, importantly, they allow limits to be set on the energy input from the lobes.

In this paper we report the initial results of an inverse Compton study of FRII radio lobes in which we look at the radio lobe dynamics, look for evidence of energy being input into the ICM, and see whether the ICM in its turn has an effect on the development of the lobes. In \citet{ine13,ine15} we modelled the large-scale environments of two samples of radio-loud AGN at redshifts $\sim$0.5 and $\sim$0.1, and here we use these models to calculate the ICM pressures around the radio lobes of the FRII galaxies in those samples. We then use radio and X-ray observations to measure the synchrotron and IC fluxes in the lobes. These allow estimates of internal lobe pressures and lobe tip Mach numbers to be made for a large, representative sample of FRII lobes for the first time. 

Throughout this paper we use a cosmology in which $H_0 = 70$ km s$^{-1}$ Mpc$^{-1}$, $\Omega_{\rm m} = 0.3$ and $\Omega_\Lambda = 0.7$. Unless otherwise stated, errors are quoted at the 1$\sigma$ level.

\section{Sample}\label{sec:sample}

The sample consists of the lobes of the FRII galaxies in the samples of radio-loud AGN at redshifts $\sim$0.5 and $\sim$0.1 used in \citet{ine13,ine15}. Table~\ref{tab:rawdata} lists the FRII galaxies used in this study, and their basic properties. The tables are arranged with the $z\sim$0.1 subsample preceding the $z\sim$0.5 subsample, and within these subsamples the sources in the individual radio surveys are listed in order of RA.

The two subsamples are as follows:

\begin{itemize}
\item The ERA sample ($0.4<z<0.6$) is a subset of the sample used by \citet{mcl04}, which was taken from the 3CRR \citep{lai83}, 6CE \citep{eal97,raw01}, 7CRS \citep{lac99,wil03}, and TexOx-1000 \citep{hil03} radio surveys. It contains 21 FRII sources ranging from $3.5\times10^{25}$ to $4.8\times10^{27}$~W~Hz$^{-1}$~sr$^{-1}$ in radio luminosity. Of these, 15 are High Excitation radio Galaxies (HERGs) and 6 are Low Excitation radio Galaxies (LERGs).

\item The z0.1 sample ($0.01<z<0.2$) was taken from the 3CRR survey \citep{lai83} and the subsample of the 2Jy survey (\citealt{wal85,tad93}) defined by \citet{dic08}. It contains 32 FRII sources (23 HERGs and 9 LERGs), with radio luminosities between $1.2\times10^{25}$ and $3.3\times10^{26}$~W~Hz$^{-1}$~sr$^{-1}$. 

\end{itemize}

A number of lobes were excluded from the analysis for practical reasons: because the lobe images were incomplete -- partially off the chip or lying across chip boundaries; because they were small in angular extent and masked by nuclear emission or by a rich ICM; because the ICM was strong and highly disturbed so the results were highly dependent on choice of background; or because the map was of poor quality or low resolution and the lobe shapes could not be defined with sufficient accuracy. In one case the lobes were so large that they extended well beyond both the maximum detected ICM radius and the $R_{200}$ overdensity radius, making estimates of the ICM pressure problematic. Table~\ref{tab:excl} lists the lobes that were excluded, and the possibility of biases in our conclusions as a result of these exclusions are discussed in Section~\ref{sec:excl} below. This left a total sample of 37 FRII galaxies, 14 from the ERA sample (11 HERGs, 3 LERGs) and 23 from the z0.1 sample (19 HERGs, 4 LERGs).

One of the sources excluded because it lies in a disturbed environment, PKS~2211$-$17 (3C~444), has a surface brightness drop around the lobes. A detailed study of the source \citep{cro11} shows cavities and a temperature drop corresponding to a Mach number of $\sim1.7$. We have included this source in the section where we consider Mach numbers (Sections~\ref{sec:M2}) but not in any analysis based on lobe internal conditions.

PKS~0625$-$53 was classified as an FRII galaxy in \citet{ine15}. It was originally classified as having a Wide Angle Tail morphology \citep{mor99}, and consequently is likely to be an FRI galaxy, but was later listed as an FRII \citep{dic08}. We have followed Mingo et al. (in preparation) and reverted to the FRI classification and so PKS~0625$-$53 has not been included in the current sample.

Where possible, we analysed the two lobes from each source separately. However, as discussed in Section~\ref{sec:shapes}, this was not always possible so for some sources we combined the lobes. The total number of lobes in the sample was 47. 14 were from the ERA sample (11 HERGs, 3 LERGs) and 33 were from the z0.1 sample (26 HERGs, 7 LERGs).

\section{Observations and data preparation}

\subsection{X-ray and radio data}
We used X-ray observations to measure the flux produced by IC in the radio lobes. The observations came from {\it Chandra} and {\it XMM-Newton} and were processed during previous studies \citep{ine13,ine15}. Observation details are listed in Table~\ref{tab:obsdata}.

We used radio maps to define the shapes of the radio lobes, so that we could measure the radio and X-ray flux from the synchrotron and IC emission and estimate lobe volume. Again, these maps were used by \citet{ine13,ine15}, and Table~\ref{tab:obsdata} contains details of the radio maps, including references.

\section{Analysis}

The aim of the analysis was to investigate the lobe dynamics and particle content by comparing the observed and minimum (equipartition) lobe pressures, and then to use these results to compare conditions within the radio lobes with those of the surrounding ICM. We needed therefore to find the internal lobe pressures, compare them with ICM pressures, and estimate a lower limit on the advance speeds of the lobe tips.

To find the internal lobe conditions, we needed to estimate the lobe volume, radio flux (from the synchrotron emission) and X-ray flux (from the IC emission).

\subsection{Lobe shapes and ICM pressures}\label{sec:shapes}

We used the radio maps listed in Table~\ref{tab:obsdata} to find the positions of the lobe tips, the mid-point of the lobes and estimate the lobe shapes and volumes. We excluded the nucleus, jets and hot-spots (which will have different electron densities and magnetic field strengths from the lobes) and background X-ray sources. Where possible, we made separate flux measurements and volume estimates for the individual lobes. For the ERA sample, however, the X-ray flux was generally very low so we used fluxes for the combined lobes. Also, for some sources there was no clear division between the two lobes, so again we combined the lobes. Details of the lobes are listed in Table~\ref{tab:lobes}.

As can be seen from Table~\ref{tab:obsdata}, we did not have a consistent set of radio maps for the sources in the samples. When possible, we used maps at the lowest available frequency to define the lobe shapes, volumes and excluded regions. However, if the low-frequency maps were of low resolution, we used higher frequency maps to define the shapes, but checked with the low resolution maps to see if there were any regions of extended flux missing from the higher frequency maps.

We then calculated the ICM pressures. In \citet{ine13,ine15}, we fitted the ICM surface brightness profiles with either single $\beta$ models (e.g. \citealt{cav76,jon84}) or, when the host galaxy extended beyond the PSF from the nucleus, double $\beta$ models which take a line-of-sight projection of the single $\beta$ models fitted to the two components \citep{cro08a}. Here we used those models to find the electron density at the lobe tips and mid-points. Pressure was then obtained following the methods of \citet{bir93} and \citet{cro08a}, by converting the electron density to gas density and applying the ideal gas law -- for this we used the ICM temperatures from \citet{ine13,ine15} (Table~\ref{tab:lobes}). The ICM mid-lobe and lobe tip pressures are listed in Table~\ref{tab:lobes}.

These pressures are dependent on the ICM temperature measurements and the $\beta$-model fits.  Since pressure scales with temperature, errors in the temperature would produce proportionally similar errors in the pressure. The temperatures were obtained where possible by spectral analysis. For the z0.1 sample, these have 1$\sigma$ errors that are mostly less than 20~per~cent of the calculated temperature, with the highest being $\sim50$~per~cent. The errors for the ERA sample tend to be larger, with 5 sources at $\sim50-70$~per~cent and one source (3C~427.1) with an upper error of $\sim1.6$ times the calculated temperature. For sources where we could not obtain a temperature from the spectrum, we estimated the temperature using the temperature-luminosity scaling relation of \citet{pra09}. When the spectral temperature was known, the estimate was usually compatible with it so the estimated temperatures are unlikely to be much in error. 

For the majority of the sources, the $\beta$-models are a good fit to the surface brightness profiles. Thus if the electron density is being estimated within the maximum detected radius of the ICM then it is likely be accurate. This is the case at mid-lobe for all but 7 lobes, and 3 of those are less than 10~per~cent beyond the detected radius. 11 lobes have the lobe tip beyond the ICM detected radius (although 4 of these are less than 15~per~cent beyond the detected radius). Thus the majority of the density estimates should be estimated well from the surface density profile.

For the lobes extening beyond the maximum detected radius, the density estimates depend on the quality of the fit of the $\beta$-model extrapolation. If this is poor then the density could be considerably in error. For 9 of these lobes, the lower 1$\sigma$ error bound is close to zero. The worst case is 3C~321, which has small lobes with the mid-lobe position at nearly three times the maximum detected radius, and is also in a sparse ICM with a steep $\beta$-model which has errors on both $\beta$ and the core radius that cover most of their expected ranges. If we replace the $\beta$-model parameters for 3C~321 with the median values, which give a much less steep $\beta$-model, the density estimate is inceased by a factor of $\sim7$.

To summarise, since the majority of the lobes in our sample lie within the maximum detected radius of the ICM, the errors introduced by uncertainties in the ICM temperature and density should be small. However, ICM pressure estimates for the few lobes lying well beyond the detected radius could be substantially in error.

\subsection{Radio flux}

We measured the radio flux density within the defined lobe shape with the nucleus, jets and hot-spots excluded. For the background, we selected a region of similar length to the lobe in an area of the map close to the lobe but outside the lobe emission.

Since the the commonest map frequency was 1.4~GHz, we used maps at this frequency when available to obtain the radio flux density. When we only had higher frequency maps, we used published flux density measurements at 408~MHz from the Molonglo Reference Catalog of Radio Sources\footnote{http://vizier.cfa.harvard.edu/viz-bin/VizieR?-source=VIII/16} or at 178~MHz from the online 3CRR catalogue\footnote{http://3crr.extragalactic.info/}. Nuclear and hot-spot emission has not been excluded from these measurements, but the flux measurements at frequencies below $\sim500$~MHz are expected to be dominated by lobe emission for the LERGs and NLRGs. We split the low flux density measurements between the lobes in the same ratio as the flux densities from the higher frequency measurements as was done by \citet{cro05b}. The frequencies used for the radio flux density measurements are given in Table~\ref{tab:flux}.

\subsection{X-ray flux}

To obtain the X-ray flux densities, we generated spectra for the lobes using \textsc{specextract}. For the background, we needed to sample regions of the X-ray background and ICM representative of the lobe region, and in some cases, the wings of the AGN PSF. Where possible, we used a rectangle of the same length as the lobe, starting at the same distance from the nucleus as the lobe and extending radially out from the nucleus. This method was suitable for well-defined lobes indented near the nucleus. For small lobes and lobes where the emission spread about the nucleus, we modelled the background using an ellipse surrounding the lobe.

We modelled the IC emission in \textsc{xspec}, using a power law coreected for Galactic absorption (\textsc{wabs(power))}. We also tried to fit a thermal model (\textsc{apec}) in case there was a contribution from the shocked ICM, but this only gave an improved fit for one source -- 3C~452. This source has been studied in detail by \citet{she11a} using an \textit{XMM-Newton} observation; we obtained compatible results. We then obtained the flux density at 1~keV by refitting the model using the \textsc{cflux} component convolved with the power law model.

We assumed for all sources that the electron distribution followed a power law with an electron energy index $\delta=2.4$ (spectral index $\alpha=0.7$) and minimum and maximum energies defined by the Lorentz factors $\gamma_{\rm min}=10$ and $\gamma_{\rm max}=10^5$. These assumptions are discussed in Sections~\ref{sec:delta} and \ref{sec:gamma}.

For some of the sources with low counts, we could get an acceptable fit from either a power law or a thermal model, but the temperature from the thermal fit was always low -- usually less than the ICM temperature -- so was unlikely to be a contribution from shocked gas. In these cases we used the power law fit. It is possible that there could be a thermal contribution from the ICM due to incomplete background subtraction, particularly for the sources with lobe emission around the nucleus. Neglecting this could result in the power law normalisation being too high, leading to high flux and internal pressure estimates. We therefore looked at the quality of the background subtraction for a selection of sources with good power law fits, and found only two sources for which we could fit a low temperature background contribution -- one of these lowered the flux by slightly more than the $1\sigma$ error; the other had a much smaller effect. This suggests that our background modelling was in general good and errors in the modelling should not have much effect on the low count sources.

For many sources, there were insufficient counts to get a good fit for the power law. For these we followed \citet{cro05b} in assuming a photon index of 1.5. If the photon index from the fit was much different from this we fixed it at 1.5 and fitted the model to obtain the normalisation. If there were insufficient counts to fit the model, we used an unbinned spectrum with the photon index again fixed to 1.5. Finally, if the net counts in the lobe were less than three times the background error we used three times the background error as an upper limit on the counts and again used a photon index of 1.5 to obtain the normalisation. Table~\ref{tab:flux} contains the radio and X-ray flux densities for the sources, together with details of the power law fits for the X-ray emission.

In all, we obtained a good fit for the photon index for 23 of the 47 lobes, and obtained a normalisation with an index of 1.5 for 18 lobes. We did not detect X-ray emission in the remaining 6 lobes.

\subsection{Internal lobe conditions}

We used the radio and X-ray flux densities, together with the lobe volumes, to obtain the equipartition and observed (inverse Compton) magnetic fields and electron energy densities. For this we followed the method described by \citet{cro05b}, using the \textsc{synch} code \citep{har98a} to find the magnetic and electron energy densities within the lobes. \textsc{synch} models the population of relativistic electrons expected from the input radio spectrum, calculating the equipartition conditions and the IC X-ray flux expected from the synchrotron and CMB photon fields. Having defined the equipartition conditions, the user can then modify the lobe magnetic field, iterating until the predicted X-ray flux matches the observed flux within the lobe. The electron density from \textsc{synch} then gives the magnetic field strength and internal lobe pressure. The model assumes that the particle content of the lobes is dominated by relativistic electrons: this assumption is discussed in Section~\ref{sec:eqp}.

Table~\ref{tab:BandP} contains the equipartition and observed magnetic fields and pressures within the lobes. The quoted errors on the observed field and pressure are derived from the X-ray flux density error, so do not include the systematic errors discussed below (Section~\ref{sec:errors}).

\subsection{Lobe-tip Mach number}\label{sec:M1}

We then used a comparison of the internal and external lobe pressures at the tip to estimate the lobe tip Mach number $\mathscr{M}$ using the Rankine-Hugoniot conditions, which are derived from the conservation of mass, momentum and energy across the discontinuity (e.g. \citealt{wor06}). Using the equation for the speed of sound in an ideal gas, assuming equal pressure throughout the lobe interior, and assuming that the shocked shell surrounding the lobe is in pressure balance with the lobe so that the internal lobe pressure can be used as a proxy for the shocked gas pressure, this gives

\begin{equation}\label{eqn:M}
\mathscr{M}^{2}=\frac{1}{2\Gamma}\left((\Gamma+1)\frac{P_{\rm i}}{P_{\rm o}}-(1-\Gamma)\right)
\end{equation}

\noindent where $\Gamma$ is the adiabatic index (5/3 for a monatomic gas), and $P_{\rm i}$ and $P_{\rm o}$ are the pressures inside and outside the lobe. As noted above, the jet ram pressure is not included in the calculation so $\mathscr{M}$ will be a lower limit -- this is discussed in Section~\ref{sec:ram}. In addition, the lobe internal pressure calculations assume that there are no non-radiating particles (e.g. protons) in the lobes. If there were a significant proton population, this would also increase the internal pressure and consequently the Mach number.

The Mach numbers are included in Table~\ref{tab:BandP}. The errors were obtained by propogating the internal and external lobe pressure errors, which in their turn come from the X-ray flux density errors and the Bayesian estimates of the ICM $\beta$-model parameters \citep{ine15} respectively. 

We have excluded sources with upper limits for the external ICM pressure $P_{\rm o}$. Note that because nearly two-thirds of the low luminosity LERGs in our original samples are FRI sources, we have Mach numbers for only 6 LERG sources (9 lobes, including 2 with upper limits for $P_{\rm i}$).

\subsection{Jet power}

Having obtained the Mach number, we calculated the speed of sound in the ICM using $c_{\rm s}=\sqrt{\frac{\Gamma kT}{m}}$ where $k$ is the Boltzman constant, and the particle mass $m$ is 0.6 times the mass of the hydrogen atom \citep{wor06}. This then gives the advance speed of the lobe and, from the length of the jet, an estimate of $\tau$, the age of the jet. Assuming the work done in displacing the ICM is similar to the internal energy of the lobe \citep{hak13,hak14}, the time-averaged jet power is then $Q_{\rm jet}\sim2E_{\rm i}/\tau$ where $E_{\rm i}$ is the internal energy density of the lobe. Jet powers are listed in Table~\ref{tab:BandP} -- the value for each source is the sum of the individual jet powers for that source. The errors were obtained by propagating the internal lobe pressure and the Mach number errors. 

The lobe tip Mach number varies during the lifetime of the radio jets -- this is described briefly in Section~\ref{sec:M2} and in more detail in, for example, \citet{hak13} -- so the timescales calculated here are estimates. More accurate estimates could be obtained by taking into account how the Mach numbers vary during the lifetime of the source.

\section{Results and discussion}

\subsection{Deviation from equipartition}\label{sec:eqp}

\begin{figure*}
  \begin{minipage}{8cm}
  \includegraphics[width=7cm]{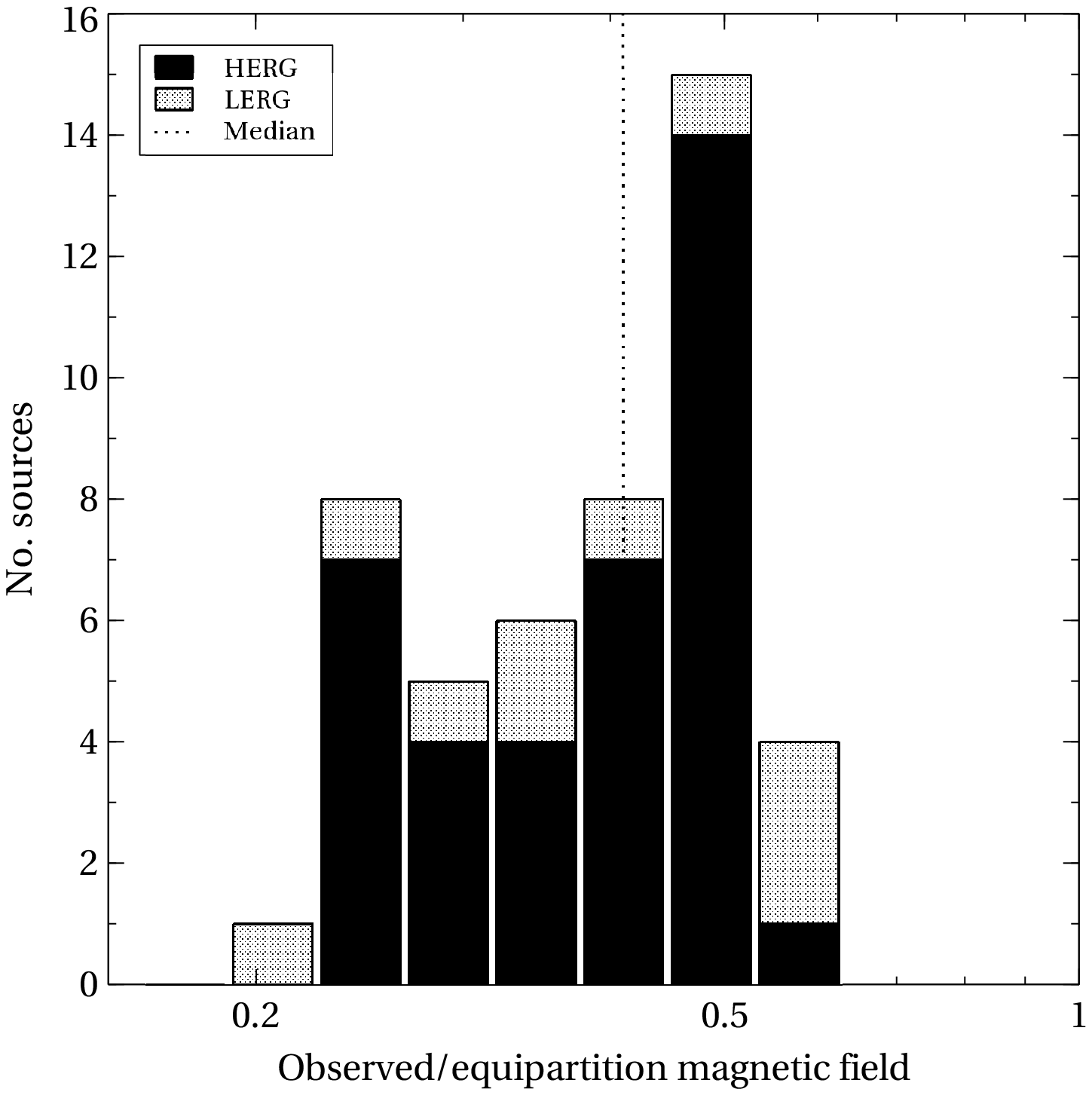}
  \end{minipage}
  \begin{minipage}{8cm} 
  \includegraphics[width=7cm]{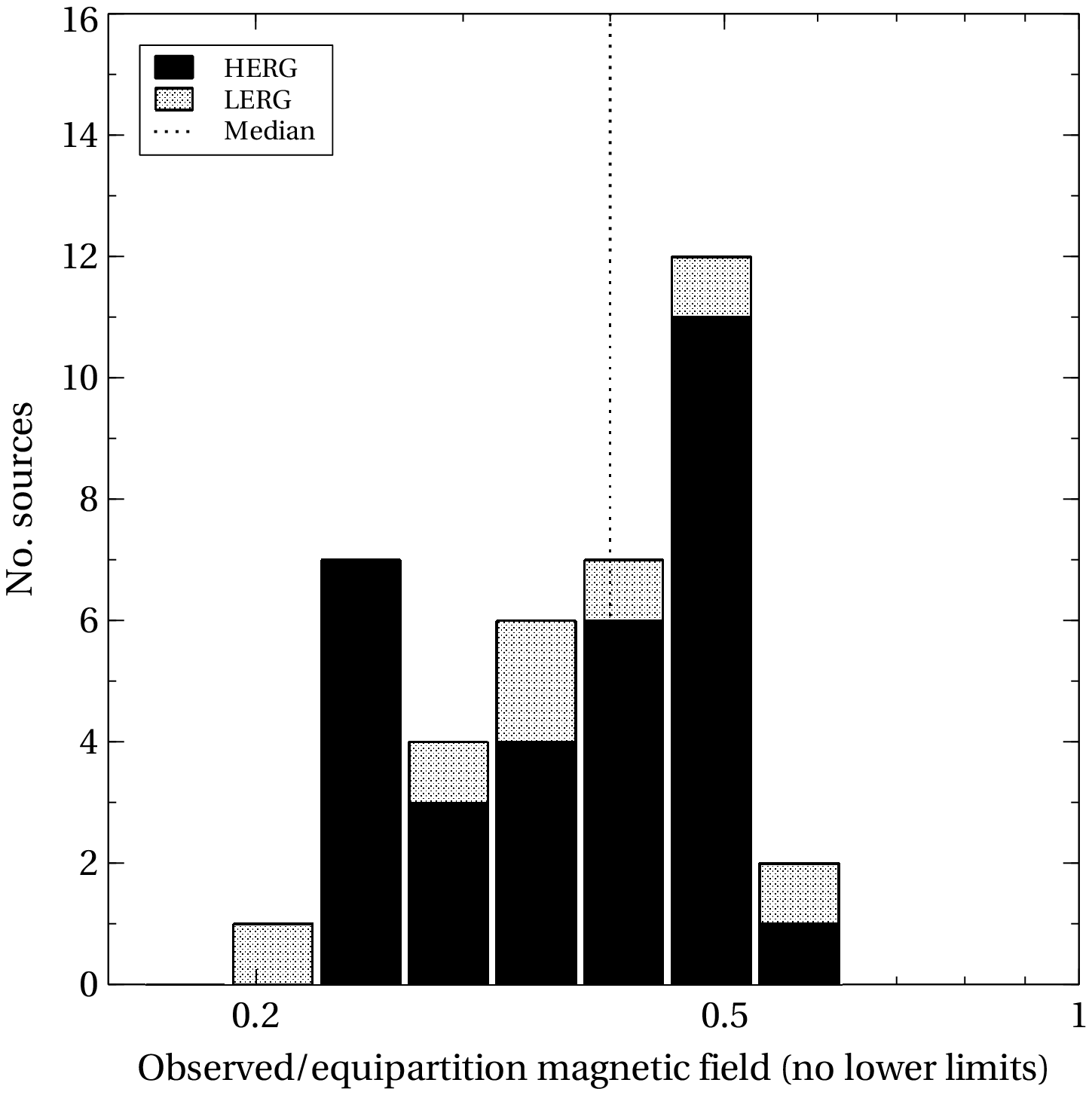}
  \end{minipage}
  \caption{The ratio of observed (calculated from the IC emission) to equipartition magnetic field strengths, separated into excitation classes. HERGs are shown in solid black, LERGs shaded. The dotted line shows the median ratio. The left histogram includes all sources; the right histogram excludes lower limits.}
\label{fig:BoBeq}\end{figure*}

\begin{figure*}
  \begin{minipage}{8cm}
  \includegraphics[width=7cm]{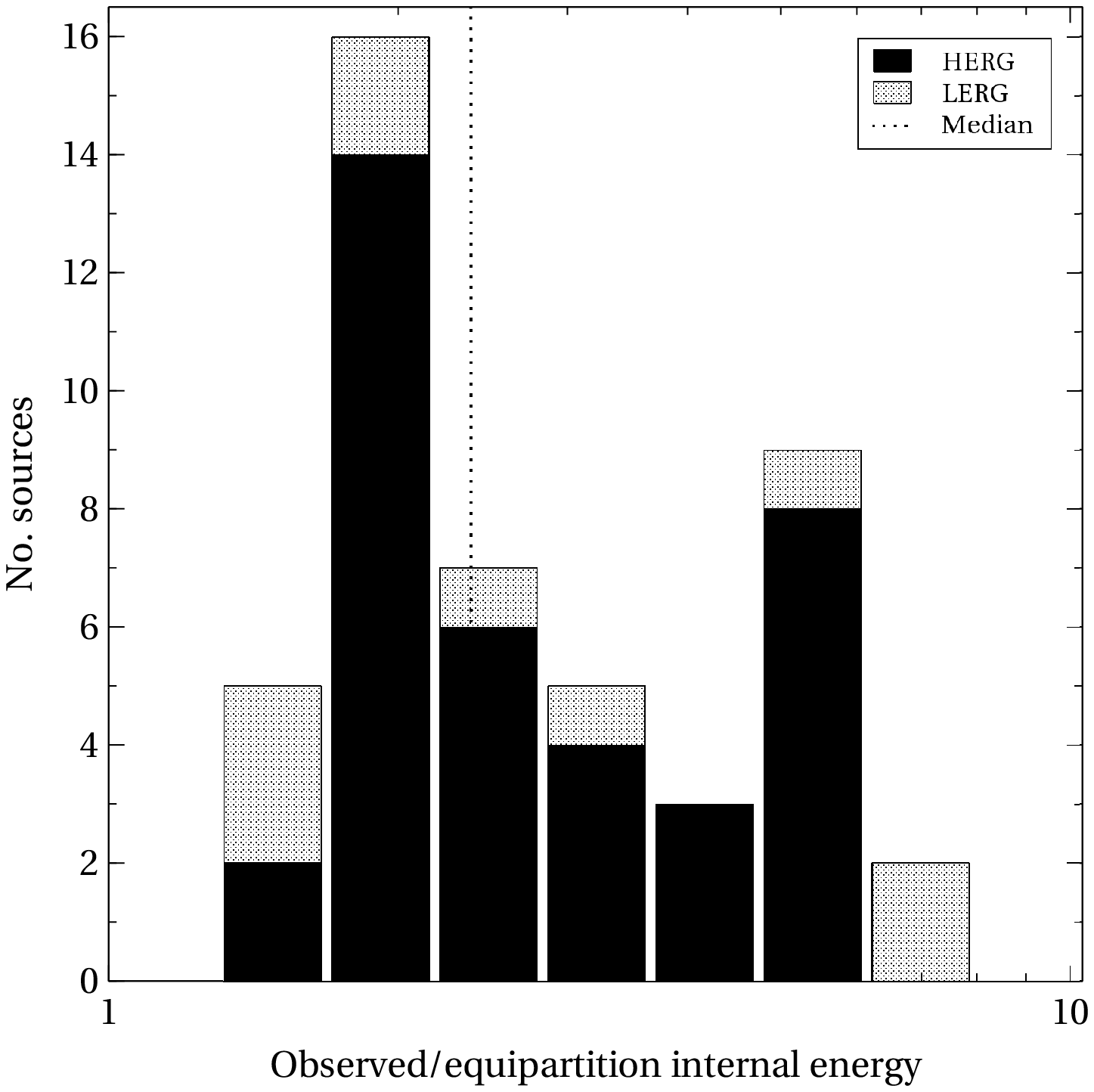}
  \end{minipage}
  \begin{minipage}{8cm} 
  \includegraphics[width=7cm]{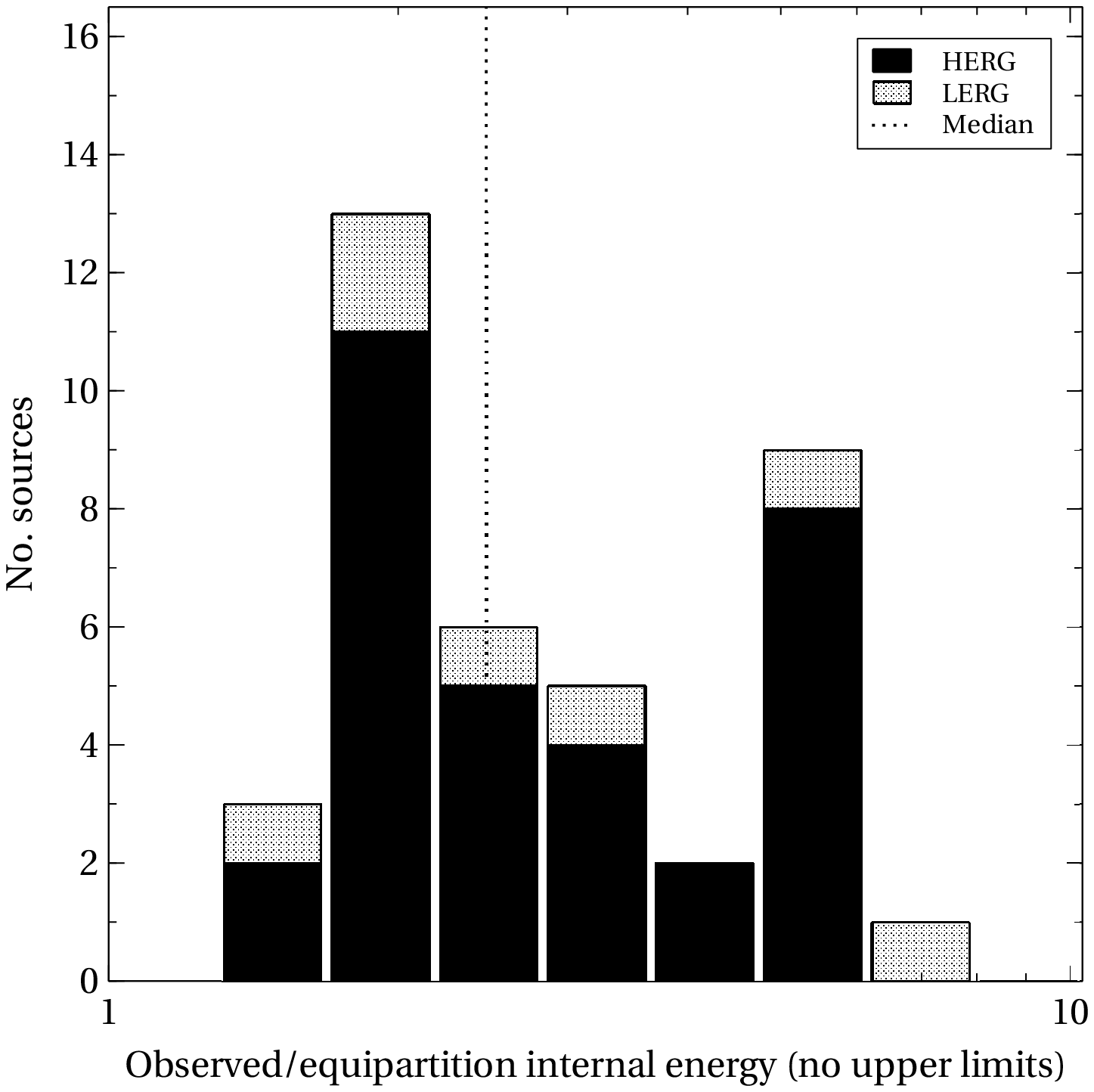}
  \end{minipage}
  \caption{The ratio of observed (calculated from the IC emission) to equipartition internal energies for the lobes, separated into excitation classes. HERGs are shown in black, LERGs shaded. The dotted line shows the median ratio. The left histogram includes all sources; the right histogram excludes upper limits.}
\label{fig:PoPeq}\end{figure*}

Figures~\ref{fig:BoBeq} and \ref{fig:PoPeq} show the ratio between the observed (IC) and equipartition measurements of the magnetic field strengths and of the total internal energies for the sample, with and without the upper limits. The observed magnetic field strength is lower than the equipartition field strength for all sources, suggesting that the lobes contain electron energy densities greater than would be implied by the minimum energy condition. All lobes are however within one order of magnitude of equipartition in terms of field strength with a median ratio of observed to equipartition fields of 0.4. \citet{har02b} and \citet{cro05b} found similar results; we have now confirmed these with a larger, more representative sample.

Similarly, the observed internal energy is higher than the equipartition internal energy for all sources and all the observed energies are within one order of magnitude of the equipartition energies. The median of the ratio of observed to equipartition  internal energies is 2.4. As with all the calculations in this chapter, these results assume that there are no protons in the lobes and that the lobes are completely filled with synchrotron-emitting particles. As discussed by \citet{cro05b}, the relatively close agreement between the equipartition and observed magnetic fields calculated assuming the lobes are dominated by relativistic electrons suggests that the lobes are not energetically dominated by protons. If there were protons in the lobes, the observed internal energy would be higher than the values calculated here and the lobes would be more over-pressured relative to the external environment.
 
\subsection{Lobe pressure balance}\label{sec:pbal}

\begin{figure*}
  \begin{minipage}{8cm}
  \includegraphics[width=7cm]{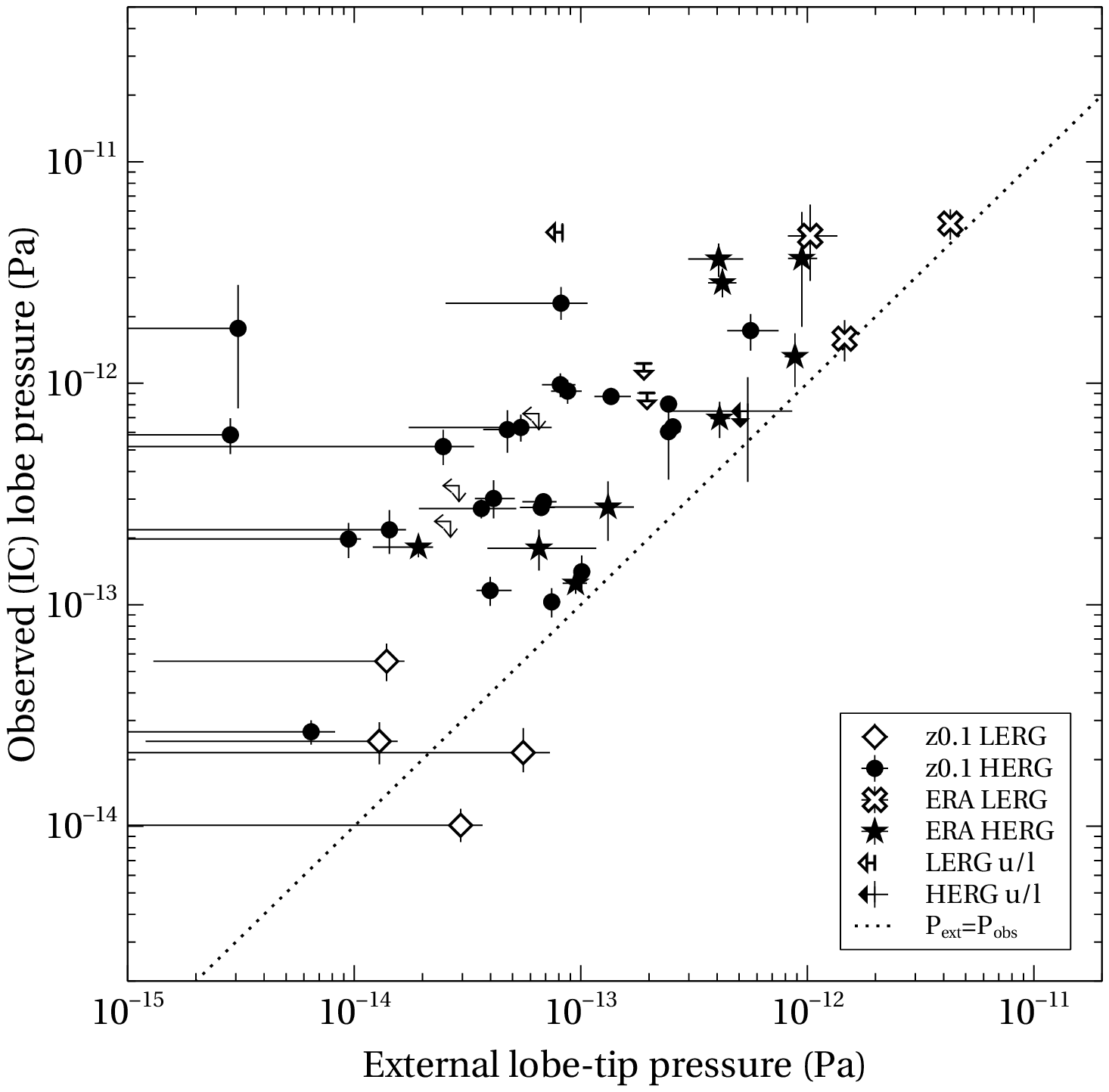}
  \end{minipage}
  \begin{minipage}{8cm} 
  \includegraphics[width=7cm]{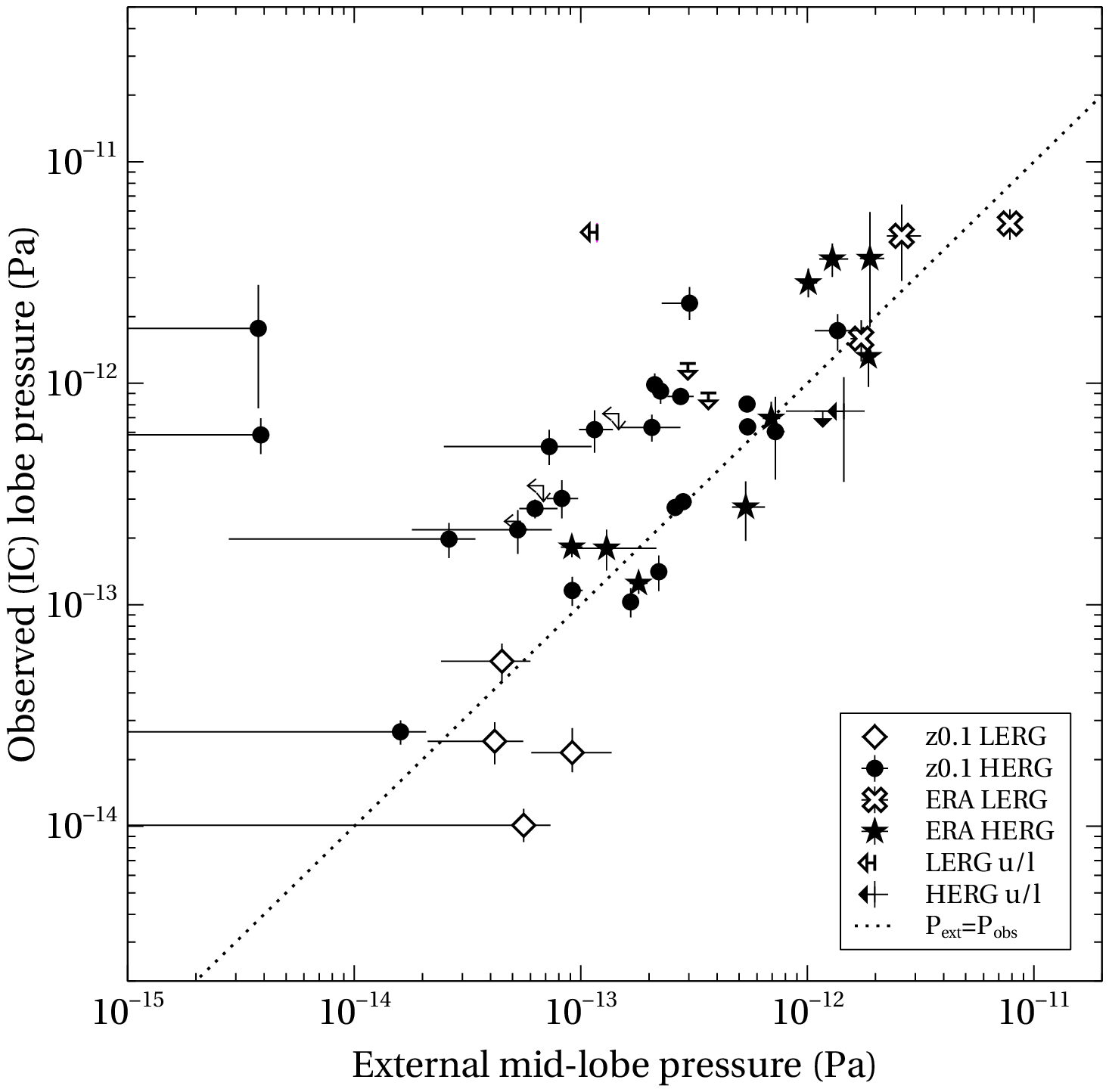}
  \end{minipage}
  \caption{Internal lobe pressure vs external (ICM) pressure, at the lobe tip (left) and at mid-lobe (right). Empty symbols are LERGs, with diamonds showing the z0.1 sample sources and crosses the ERA sources. Black symbols are HERGs with circles showing the z0.1 sources and stars the ERA sources. Arrows denote upper limits.}
\label{fig:PexPobs}\end{figure*}

Figure~\ref{fig:PexPobs} shows the observed pressure plotted against the external pressure at the lobe tip and at mid-lobe. Pressure balance is shown as a dotted line. If the lobes are growing they should be over-pressured at the tip, and this is the case for all the sources in our sample except 3C~326 (which has the largest lobes in our sample by several hundred~kpc, so the external pressure is from an extrapolated $\beta$-model). Since the effects of the jet ram pressure have not been included, the internal lobe-tip pressure will be higher than the observed lobe pressure calculated here, and so the observed pressures at the tip should be regarded as lower limits. The true tip pressure could plausibly be above pressure balance for all the sources. Our median lobe-tip pressure ratio of 4.3 lies well within the range found in the simulations of \citet{hak13}; our spread is greater but we have a wider range of environments than were used in the simulations.

At mid-lobe, the lobes are distributed about the pressure balance line with a median $P_{\rm obs}/P_{\rm ext}$ of 1.8 -- \citet{cro05b} and \citet{bel07} also found FRII lobes to be near pressure balance at mid-lobe. The lowest ratio of observed to external pressure is 0.18 for 3C~326, which also has the lowest lobe-tip pressure ratio. There are only three lobes more than one order of magnitude above pressure balance. The lower limit PKS~0034$-$01 ($P_{\rm obs}/P_{\rm ext}=41$) has small lobes (42~kpc long) so may be a young source. 3C~321, which has very high pressure ratios of well over 100, has small lobes (50 and 90 kpc long) 280~kpc from the nucleus. It is merging with a neighbouring galaxy and is in a disturbed environment \citep{eva08}. The $\beta$-model has steep, poorly constrained parameters, giving very low ICM pressures at mid-lobe and so the extrapolation of the model to obtain the external lobe pressures is likely to be imprecise. As discussed in Section~\ref{sec:shapes}, replacing the $\beta$-model parameters with the median model parameters increases the mid-lobe external pressure estimate considerably.

\citet{bel07} had only two LERGs in their sample (3C~427.1 and 3C~200), but found that both of these appeared underpressured at mid-lobe, giving the possibility that LERG lobes contain non-radiating particles such as protons to provide additional pressure. However, although we used the same observations for these sources as \citet{bel07}, we obtained lower external pressures, and although these two LERGs occupy richer environments than any of the HERGs in our sample, our pressure ratios for them do not stand out from the rest of the sample.

The lowest pressure ratio lobes in our sample are also from a LERG source (3C~326), although the pressure ratios of the other LERG sources (except the lower limit PKS~0034$-$01 mentioned above) lie within the range of results for the HERGs. The median pressure ratios are 1.9 and 0.9 for the HERGs and LERGs respectively and a median test shows no evidence for a difference between the HERG and LERG mid-lobe pressure ratios ($\chi^2=0.7$, $p=0.4$). Our pressure estimates assume that the radiating particles in the lobes are leptons; the fact that our LERG results do not stand out from the HERG results suggests that the two types of radio galaxy have similar lobe contents, and therefore that the apparent difference in FRI and FRII particle content discussed in Section 1 is related to large-scale morphology and environmental interaction rather than differences in central engine.

Our results therefore show no evidence of a systematic difference between HERG and LERG lobes, but it should be noted that, since the majority of FRII galaxies in our sample are HERGs, there are only 10 LERG lobes (including 3 upper limits) in our sample of 44 lobes.

\subsection{Lobe tip Mach number}\label{sec:M2}

\begin{figure*}
  \begin{minipage}{8cm}
  \includegraphics[width=7cm]{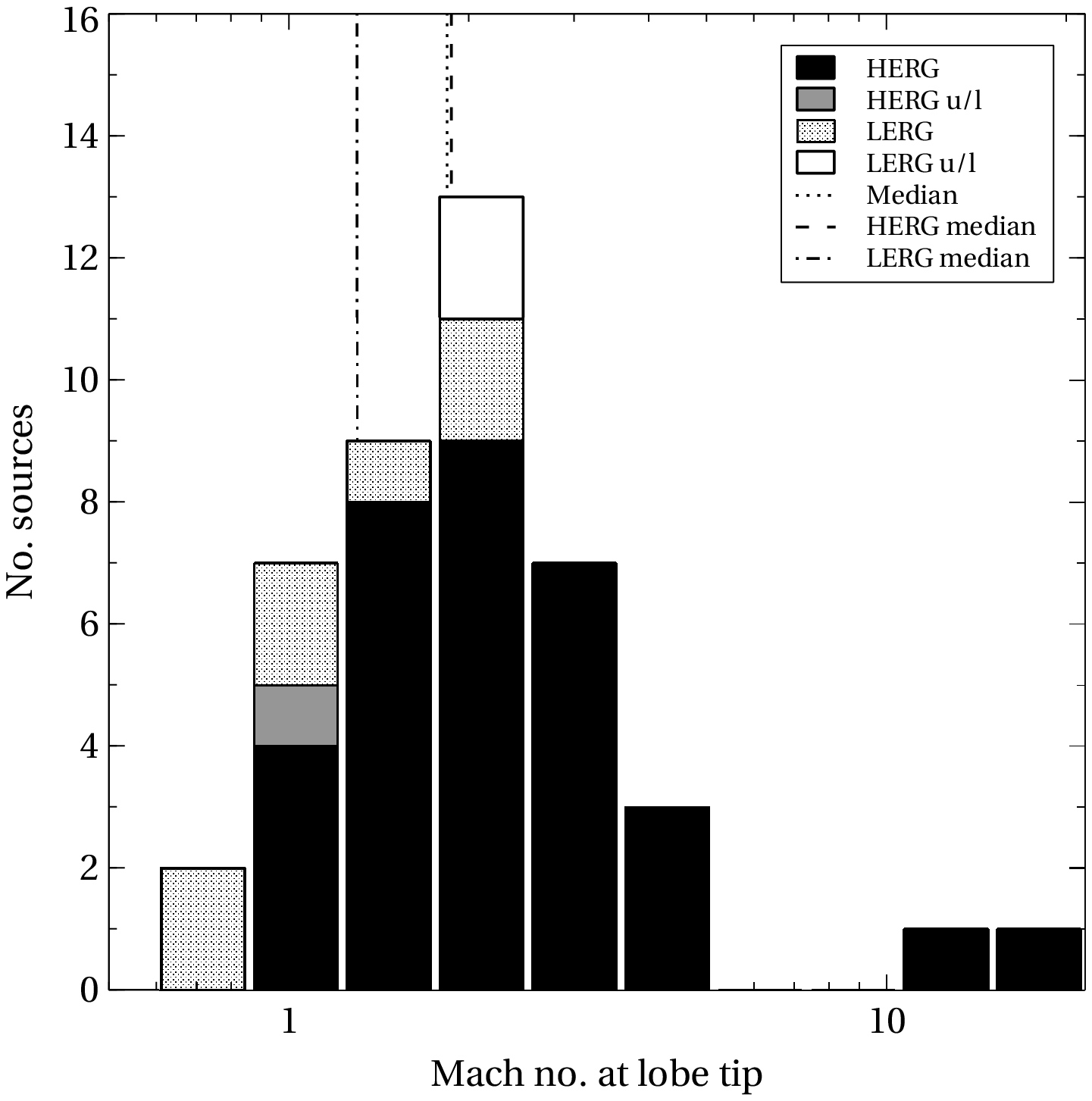}
  \end{minipage}
  \begin{minipage}{8cm}
  \includegraphics[width=7cm]{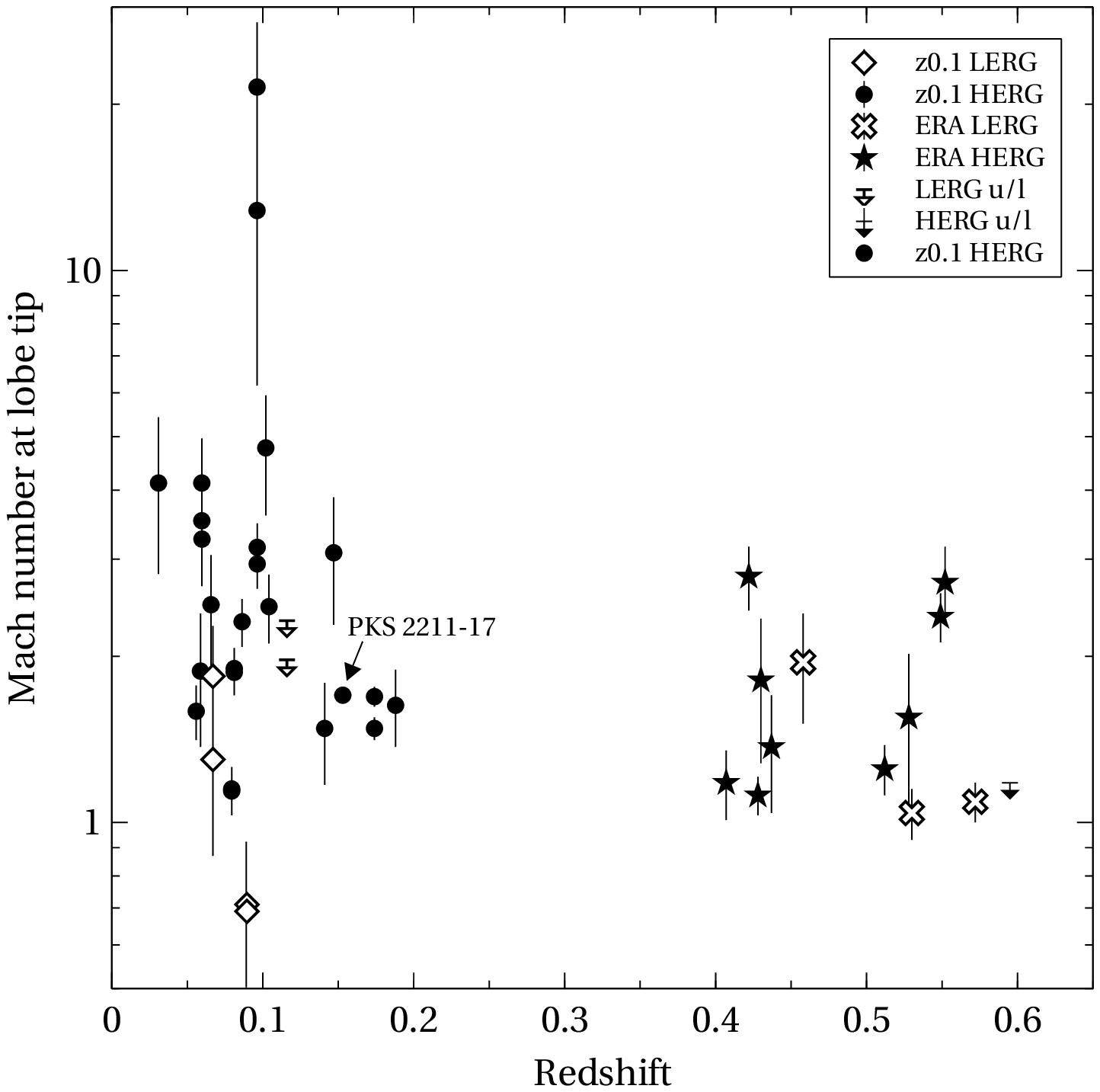}
  \end{minipage}
  \caption{On the left, Mach number at the lobe tip. HERGs are shown in solid black and LERGs shaded, HERG upper limits are solid grey, LERG upper limits are white. The HERG and LERG median Mach numbers are shown by dashed and dash-dotted lines respectively, and the overall median by the dotted line. On the right, Mach number at the lobe tip plotted against redshift. Symbols as in Figure~\ref{fig:PexPobs}. The two Mach numbers greater than 10 are from the lobes of 3C~321, discussed in Section~\ref{sec:M2}}
\label{fig:Mach}\end{figure*}

The Mach numbers are plotted in Figure~\ref{fig:Mach}. As mentioned in Section~\ref{sec:M1}, these Mach numbers neglect the jet ram pressure. However, we discuss the possible magnitude of the ram pressure in Section~\ref{sec:ram} and conclude that it is likely to be small compared with the lobe internal pressure. We therefore regard these Mach numbers as a reasonable estimate of the true value. Our results for 3C~452 are similar to those of \citet{she11a}, who obtained a Mach number of 1.6 for the combined lobes of that source.

The two lobes with Mach numbers greater than 10 are both from 3C~321, which, as mentioned above, has very low, uncertain ICM pressures. Replacing the ICM pressures with pressures estimated using the median $\beta$-model parameters reduces the Mach numbers of the two lobes to $\sim5$ and $\sim7$ -- still high but not unreasonable. We have not included 3C~321 in any further plots, but all statistics have been calculated with and without 3C~321.

Without 3C~321, the maximum Mach number is 4.8 and the median is 1.8, so the results of our large sample lie within those of the individual and small-sample studies listed in Section~\ref{sec:intro} -- of the eight examples listed there, only two sources have lobes advancing at more than Mach 2. Our median also lies within the range of simulation results of \citet{per14} for well developed lobes, supporting their choices of conditions.

\subsubsection{Magnitude of the Mach number}

The majority of the examples in Section~\ref{sec:intro} are FRI galaxies, with two of mixed morphology and only one pure FRII. It might be expected that the more powerful FRII galaxies should have higher Mach numbers, but both backflowing plasma and changes in jet direction would disperse the jet momentum over a wider region than the jet radius and would reduce the Mach number from that expected for a static jet. Evidence that the latter process is commonplace is found in observations of multiple hot-spots and of offsets between the hot-spots at different frequencies (e.g. \citealt{lai81,har07a}), and Mingo et al. (in preparation), looking at the extended emission in the 2Jy sources, found several examples in the sources used in this study.

When comparing the properties of FRI and FRII radio galaxies, it also needs to be remembered that, as mentioned in Section~\ref{sec:intro}, FRI galaxies are thought to entrain material from the ICM with the amount of entrainment being related to the richness of the environment. This introduces a non-radiating proton population, increasing the internal pressure of the jet and resulting in an FRI galaxy having a higher jet power than an FRII galaxy of the same radio luminosity. Thus it is to be expected that FRII Mach numbers will in general be lower than those of FRIs with similar luminosities.

In addition, the expansion speed of a radio lobe will depend on its stage of evolution (e.g. \citealt{hei98,hak13}). The jet extends rapidly in its early stages before expanding sideways as the shocked material at the tip starts to spread. The Mach number is therefore expected to be high for young sources. For example, the highest Mach number in the examples listed in Section~\ref{sec:intro} (8.4, \citealt{cro09}) comes from the inner lobes of Centaurus A, which are only a few kpc long.

Taking these points into account, it is understandable that the Mach numbers for our mature FRII sample should have a similar range to those observed in local FRI galaxies.

\subsubsection{HERG/LERG subsamples}

Our HERG subsample has a wider range of Mach numbers than the LERG subsample, and the median Mach number for the LERGs is lower than that of the HERGs. However, a median test does not reject the null hypothesis that the medians are the same ($\chi^2=0.26$, $p=0.6$), and as mentioned above, this gives us no reason to suppose that the lobes of the two types of radio galaxy should have different dynamics. The difference in range of Mach numbers probably reflects the different distribution of environments \citep{ine15}.

\subsubsection{Particle content}

As stated before, these Mach numbers assume that the lobes do not contain protons. The presence of a significant proton population such as is thought to exist in FRI lobes would increase the lobe internal pressure substantially. We therefore looked at the effect of a ten-fold increase in internal pressure on our median Mach number and found that this would raise it to 5.7 -- near the top of the range of Mach numbers found in the observational studies listed in Section~\ref{sec:intro}. It is unlikely that our sample median would be so high compared with the Mach numbers calculated from observed shocks, which would suggest that any proton content is low compared with that of FRI lobes.

\subsubsection{Mach number and radio luminosity}

\begin{figure}
  \centering
  \includegraphics[width=7cm]{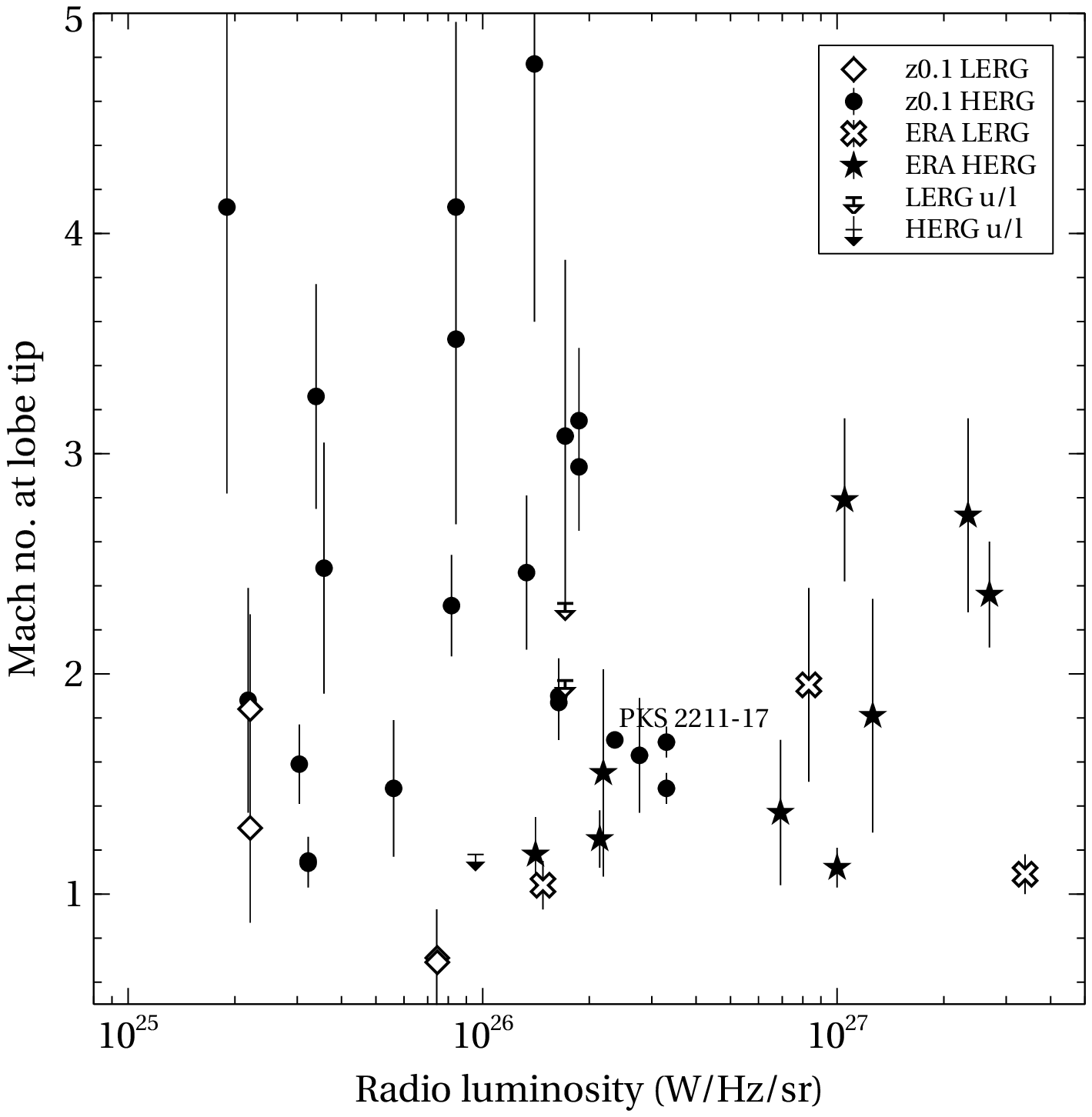}
  \caption{Radio luminosity plotted against Mach number. Symbols as in Figure~\ref{fig:PexPobs}.}
\label{fig:MLr}\end{figure}

It would be useful if the lobe dynamics, and consequently energetic input could be inferred directly from the radio and/or environment properties. As a first step we searched for any relationship between radio luminosity and Mach number (Figure~\ref{fig:MLr}). The plot shows no sign of a correlation, and this is confirmed by a generalised Kendall's $\tau$ test (Table~\ref{tab:KTau}). 

\subsubsection{Mach number and ICM richness}

\begin{figure*}
  \begin{minipage}{8cm}
  \includegraphics[width=7cm]{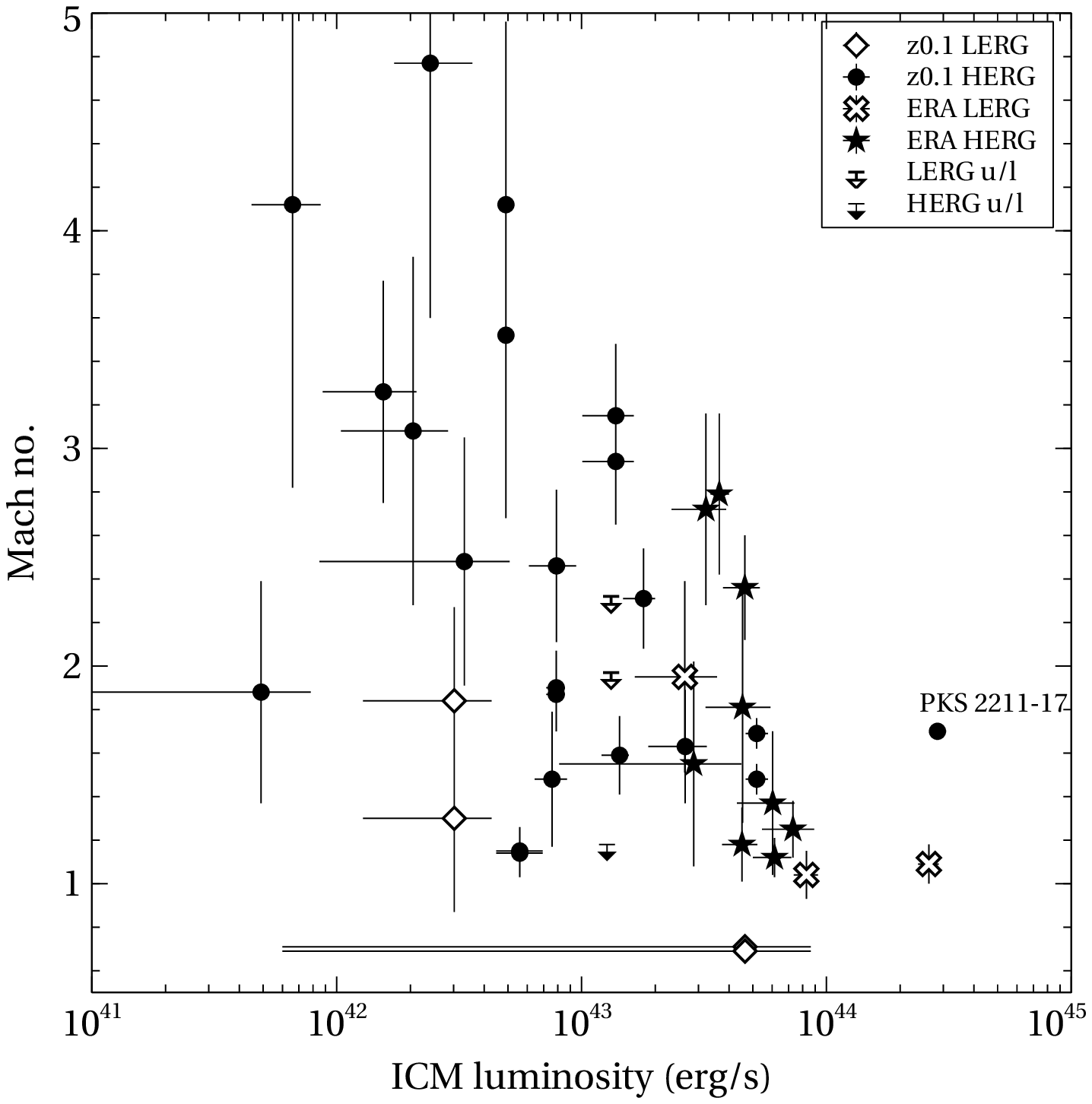}
  \end{minipage}
  \begin{minipage}{8cm} 
  \includegraphics[width=7cm]{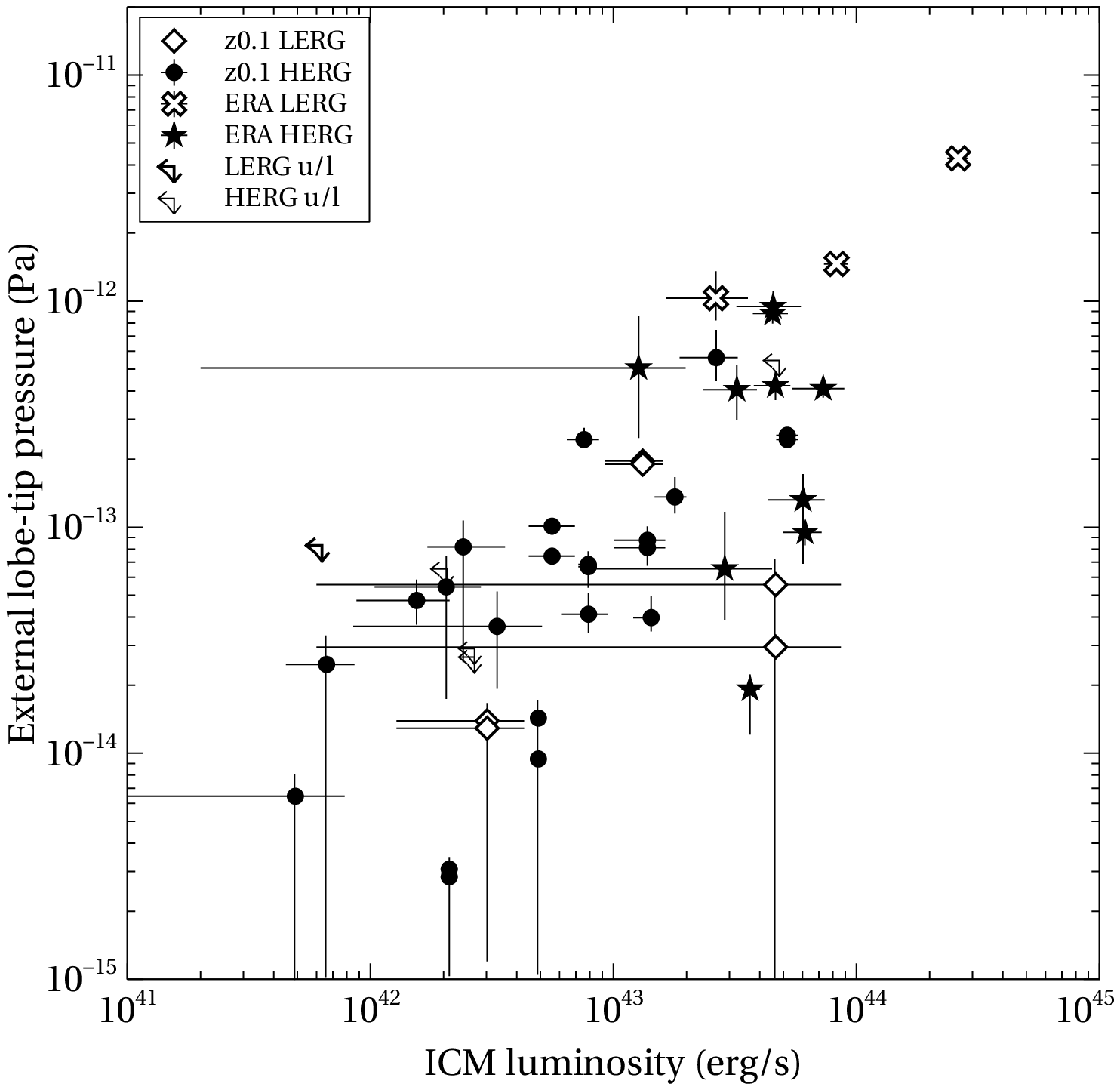}
  \end{minipage}
  \caption{ICM X-ray luminosity plotted against Mach number (left) and lobe-tip pressure (right). Symbols as in Figure~\ref{fig:PexPobs}.}
\label{fig:MLx}\end{figure*}

We then compared Mach number with ICM richness (Figure~\ref{fig:MLx}, left), using the ICM X-ray luminosities from Table~\ref{tab:lobes}. In this case Mach number appears to decline with increasing ICM luminosity for the HERG subsample, while the LERG sources are limited to a band of ICM luminosities at low Mach number. The generalised Kendall's $\tau$ tests (Table~\ref{tab:KTau}) show a strong negative correlation for both the full sample and the HERG subsample. The correlation is weakened slightly when the uncertain 3C~321 Mach numbers are removed, but is still strong. There is quite substantial scatter -- at least some of this is likely to be due to systematic errors which are discussed in Section~\ref{sec:errors} below. 

It must be noted however that ICM external pressure was used in the Mach number estimates. As can be seen from the right-hand plot in Figure~\ref{fig:MLx}, lobe-tip external pressure correlates very strongly with ICM luminosity. Since both factors have a dependence on redshift, we used a partial Kendall's $\tau$ test to confirm that the correlation was not driven by redshift (Table~\ref{tab:KTau}).

\begin{figure}
  \centering
  \includegraphics[width=7cm]{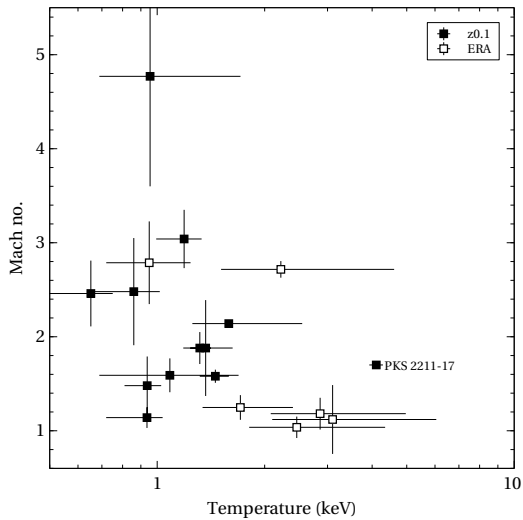}
  \caption{Mach number vs ICM temperature for the sources with temperatures calculated from the ICM spectrum. Solid squares are sources from the z0.1 sample and empty squares from the ERA sample.}
\label{fig:kTM}\end{figure}

It seems likely that lobe-tip external pressure is the factor driving the correlation between Mach number and ICM luminosity. However, ICM temperature is related to ICM luminosity and is also used in the pressure calculations so is a potential confounding factor. Temperatures were obtained wherever possible from the ICM spectrum, but when there were insufficient counts they were estimated using the luminosity-temperature scaling relation of \citet{pra09}. We therefore removed these estimated temperatures from the sample in case they biassed the results; the correlation between Mach number and luminosity was considerably weakened but was still present (Table~\ref{tab:KTau}). Then we checked to see if Mach number was related to the spectrum-derived ICM temperatures. As can be seen from Figure~\ref{fig:kTM}, there is no sign of a correlation. Thus it seems possible that there is a relationship between Mach number and environment richness, but more work is needed to eliminate potential confounding factors.

That there should be some dependence of lobe advance speed on the density of the environment is to be expected and so it seems reasonable that a lobe in a weak environment should advance more quickly than one in a rich environment. However, the environment density reduces with distance from the centre of the cluster. In the simulations discussed in \citet{hak13}, the jet initially extends rapidly and then slows as it starts forming lobes. Then as the jet tip reaches the sparser environment beyond the ICM core radius it speeds up again. Thus even if all other factors remain the same the Mach number will not be constant throughout the life of the radio galaxy. More work is therefore needed to look at how the Mach numbers of jets in environments of different richness change as the lobes grow and age, and how they are affected by other factors such as jet power.

If this negative correlation is genuine, it will have contributed to the lack of high Mach number LERGs -- since ICM luminosity correlates with radio luminosity for LERGs \citep{ine13,ine15} and the low radio luminosity LERGs mostly have FRI morphology, the LERG lobes remaining in the sample are for the most part in relatively rich environments and so would be expected to have low Mach numbers.

\subsection{Jet power}

\begin{figure*}
  \begin{minipage}{8cm}
  \includegraphics[width=7cm]{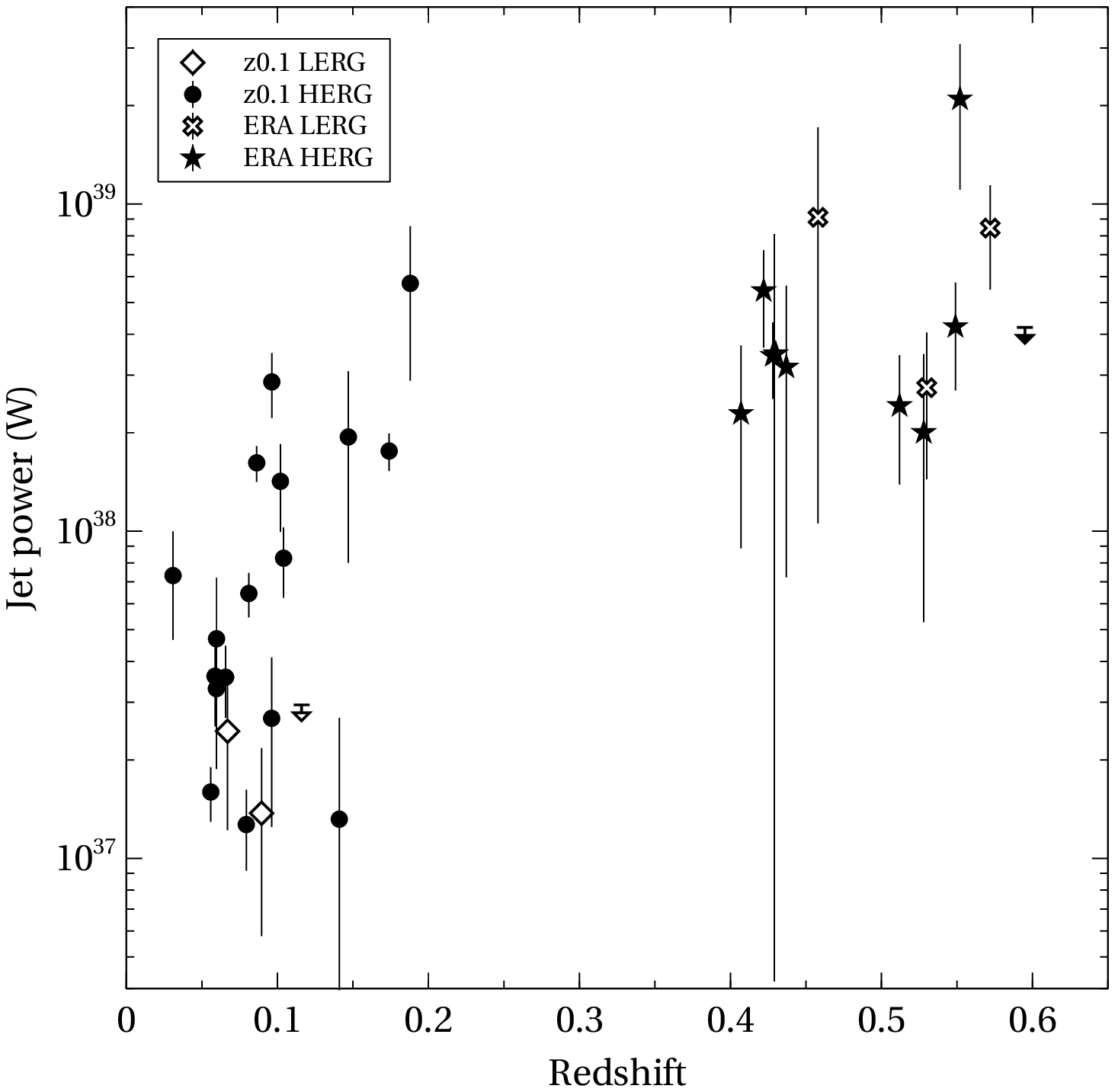}
  \end{minipage}
  \begin{minipage}{8cm} 
  \includegraphics[width=7cm]{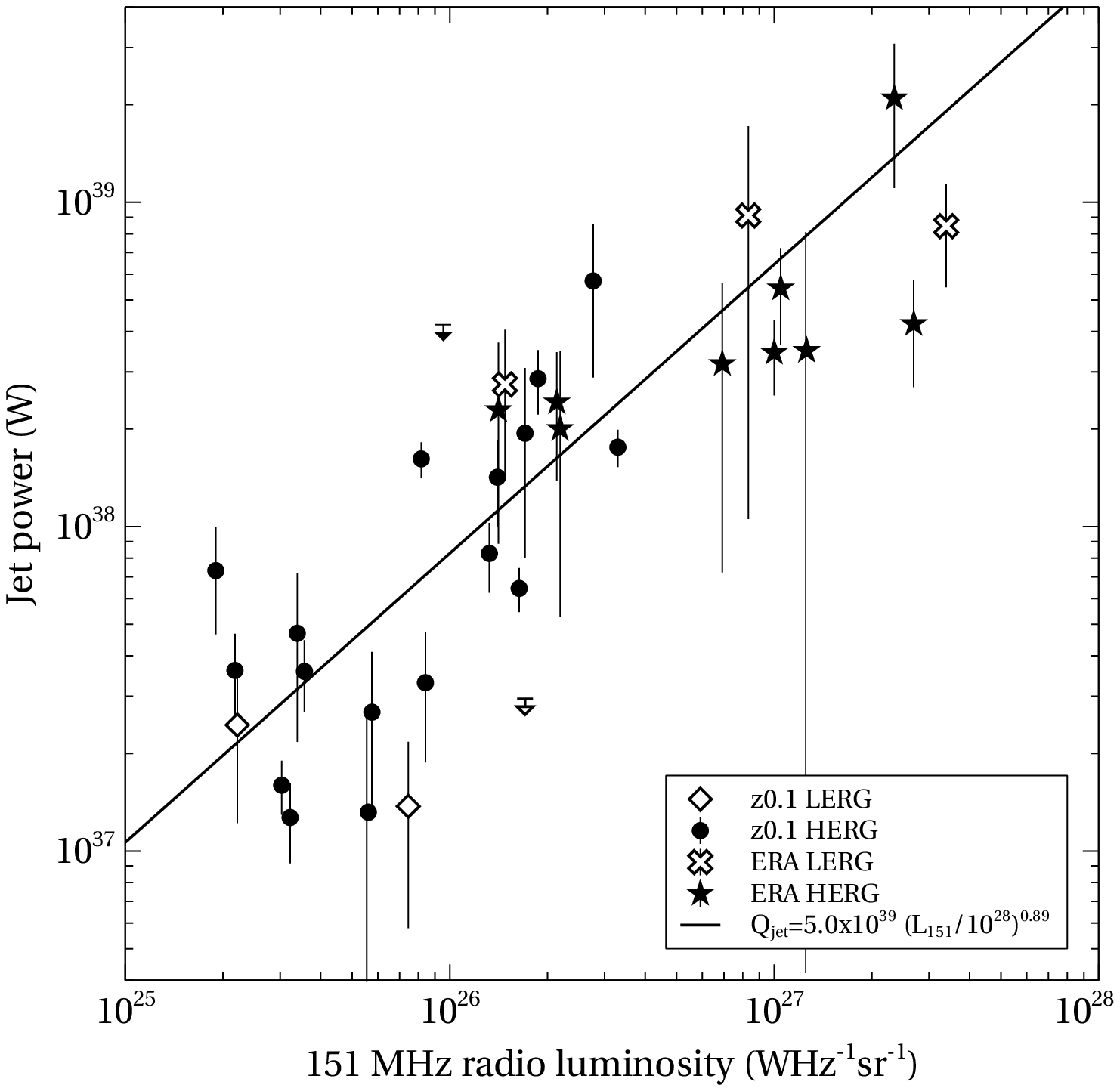}
  \end{minipage}
  \caption{Jet power plotted against redshift (left) and radio luminosity (right). The regression line was calculated using the Buckley-James method \citep{iso86}. Symbols as in Figure~\ref{fig:PexPobs}.}
\label{fig:Qjet}\end{figure*}

Figure~\ref{fig:Qjet} (left) shows jet power plotted against redshift. Since the sample is taken from flux-limited surveys, the strong redshift dependence is expected.  Our jet powers, being for FRII morphology sources, lie in and above the upper part of the ranges of FRI jet powers found by \citet{bir08},\citet{cav10} and \citet{osu11a}, and have a very similar range to the FRII sample of \citet{god13} and the lower regions of the sample of powerful FRIIs of \citet{dal12}. Of the seven sources we have in common with \citet{god13} and \citet{dal12}, we obtained similar $Q_{\rm jet}$ estimates for all but one source (3C~321 -- see Sections~\ref{sec:pbal} and \ref{sec:M2}).

In Figure~\ref{fig:Qjet} (right) we show the relationship between jet power and radio power. A correlation is expected as the two quantities are not completely independent: the jet power estimates have a complicated dependence on the measured radio flux and redshift. However, we are interested in comparing the jet power measurements and the slope and normalisation of the relation with other methods. We find a strong correlation in the presence of a common relationship with redshift ($z=4.11$, $p<0.0001$ -- Table~\ref{tab:KTau}). Since there were only two upper limits, we excluded these and calculated the regression line using the BCES regression method \citep{akb96} so that we could include the efect of the calculated errors. This gave $Q_{\rm jet}=5\times10^{39}L_{151}^{0.89\pm{0.09}}$ where $Q_{\rm jet}$ is in Watts and $L_{151}$ is the 151~MHz radio luminosity in units of $10^{28}$~W~Hz$^{-1}$~sr$^{-1}$. \citet{wil99}, assuming the FRII lobes were self-similar and at minimum energy, obtained the relation $Q_{\rm jet}=3\times10^{38}f^{3/2}L_{151}^{6/7}$ where $f$ is a factor expected to depend on the properties of the source and its environment and predicted to lie between 1 and 20. Thus our $Q_{\rm jet}-L_{151}$ relationship has a very similar slope to that obtained by Willott~et~al., but it requires $f\sim6.5$ to match the normalisation. \citet{dal12} found a factor of $f\sim4$ for their sample of FRII galaxies, and \citet{tur15} found that a factor of $f=5$ was needed to fit their dynamical model results to the theoretical model.

Since the equation derived by \citet{wil99} assumed minimum energy conditions for the lobes, which give very similar total internal energies to the equipartition conditions, we calculated the jet powers resulting from an assumption of equipartition and obtained a regression line $Q_{\rm jet}=2\times10^{39}L_{151}^{1.0\pm{0.07}}$. The difference in the normalisations of our two relations is compatible with our median ratio of observed to equipartion internal lobe energy of 2.38. The normalisation for our equipartition equation is however still higher than that of Willott~et~al. by a factor of 6.7 ($f\sim3.5$).

Another potential contribution to $f$ comes from the viewing angle of the lobes; assuming this is random gives a mean factor of $\sim1.4$ \citep{wil99} and reduces the unexplained contribution to $f$ to $\sim2.5$. 

At least some of this can be attributed to the environments of the radio galaxies. \citet{wil99} use a sample of relatively distant quasars which have relatively rich environments; using a particle density more typical of low redshift FRII galaxies reduces their estimate of the age of the system by a factor of $\sim0.5$, thus increasing the jet power by a factor of 2. \citet{hak13} report a similar dependency on environment in their simulations. Although the slope of their time/jet length relationship eventually matches the slope of the theoretical relationship used by Willot~et~al, the normalisation is dependent on the properties of the ICM. Willott~et~al. use a simple power law relationship for the change in ICM density with radius rather than the now more commonly used isothermal $\beta$-model, so their time/jet-length relationship does not include the effects of the changes in ICM density profile around the core radius. So for a source with $\beta$ similar to our median (and to the value of $\beta$ used by Willot~et~al.), Hardcastle and Krausse obtained source ages close to the theoretical relationship for their 80~kpc core radius but the age was reduced by almost one-third for a 40~kpc core radius -- the median for our sample -- which would raise the jet power by a factor of 1.5.

However, the normalisations obtained in the simulations include the effect of the initial rapid growth of the jet, which, as discussed in \citet{hak13}, is not modelled accurately, so this factor of 1.5 may not be accurate. But these simulations do confirm that the jet ages of FRII galaxies in weaker environments than those used by \citet{wil99} are likely to be shorter than their theoretical prediction and so will contribute to our higher normalisation of the $Q_{\rm jet}-L_{151}$ relationship. Also, since our core radii cover a wide range (see Table~\ref{tab:beta}), the differences in the distance that the jet travels before the ICM density drops will contribute to the scatter in our relationship.

Our estimates of jet age are based on the current Mach number of the lobe tip and so do not take into account either the early growth of the jet or the variations due to the ICM density profile. Nevertheless, our estimates of jet power from lobe internal energy and lobe-tip Mach number produce a $Q_{\rm jet}-L_{151}$ relationship that is similar to the theoretical relationship developed by Willott~et~al. once the effects of the higher electron energy densities and the different richness of the ICM environments have been taken into account.

We caution that radio luminosity is only expected, from numerical modelling, to be roughly constant with time once the source has grown to a size comparable to the host environment core radius (e.g. \citealt{hei98,hak13,eng16}). During the early stages of a source's evolution the radio luminosity must necessarily depend on source age as well as jet power. The nature of our selection means that we are biased against small, young, faint sources and so we would expect to recover a simple jet power/radio luminosity relationship in our sample.

\subsection{Lobe volumes}\label{sec:Vrad}

Finally, we compared the length and volume of the radio lobes. If the lobes are self-similar, $V\propto L^3$ where $V$ is the lobe volume and $L$ its length \citep{kai97}; if however the lobes are being partially modified and constrained by the ICM, then self-similarity would be broken \citep{hak13}.

\begin{figure}
  \centering
  \includegraphics[width=7cm]{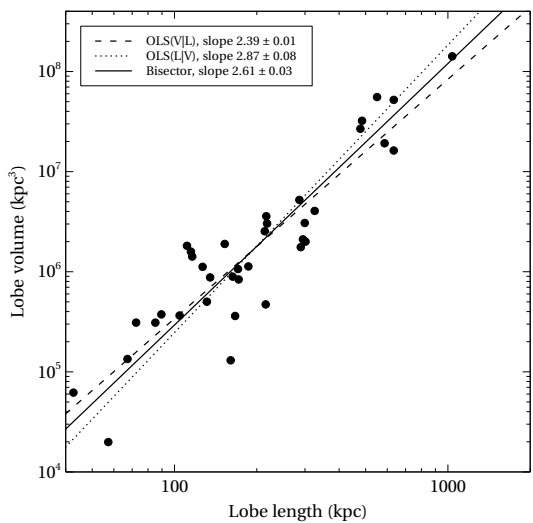}
  \caption{Lobe volume plotted against lobe length -- the mean lobe length and volume were taken for each source. Also shown are the OLS regression lines taken from both directions (dotted and dashed lines) and the bisector (solid line).}
\label{fig:Vlen}\end{figure}

The average lobe length and volume for each source are plotted in Figure~\ref{fig:Vlen}. Note that 3C~321, which has two small lobes a great distance from the nucleus, has been excluded -- the ratio of lobe length to lobe tip distance for the rest of the sample ranges from 0.18 to 0.25 whereas those for the 3C~321 lobes are 0.05 and 0.08.

There is a clear correlation between the two factors. The errors are unknown but are present in both factors, so we calculated Ordinary Least Squares (OLS) regression lines for both volume vs length and length vs volume and took the bisector \citep{iso90} -- this gives $V\propto L^{2.51\pm{0.02}}$. The three regression lines are shown in Figure~\ref{fig:Vlen}. They all have a slope of less than 3 which suggests the lobes deviate slightly from self-similarity. Departures from self-similarity have also been reported by \citet{mul08}, who found that the axial ratio reduces with increasing lobe length. This result fits with our previous results showing that the lobes are only partially over-pressured, and once again supports Model C from \citet{scu74}. It also supports the simulation results of \citet{hak13}.

\subsection{Implications}\label{sec:impl}

As found by other researchers (e.g. \citealt{har02b,iso02,cro04,iso05,kat05,cro05b}), the internal energies and pressures of the lobes are near but slightly above equipartition, reflecting the contribution of a higher electron density than required by the minimum energy conditions. If there were a sizeable population of non-radiating particles in the lobes, as typically seen in FRI radio galaxies (e.g. \citealt{cro08a}), we would expect a larger departure from equipartition (e.g. the electon energy densities in FRI lobes are typically a large factor below equipartition). The lobes were all within one decade of equipartition, which suggests both that there is not a large population of energetic protons and that the filling factor must be close to unity. Importantly, this means that (i) inverse-Compton analysis provides a reliable measure of the total internal energy for individual FRII radio galaxies, and (ii) for samples of FRII radio galaxies without X-ray data, the total energy (and magnetic field strength) can be estimated reliably from radio observations alone, by assuming the typical departure from equipartion measured here ($\sim 2.4$ in totall energy density and $\sim 0.4$ in magnetic field strength).

Our measurements of the pressure ratio between the lobes and the environment provide a test for dynamical models of FRII evolution, as well as enabling a population-wide estimate of jet energetics. The pressure difference between the lobes and their environment is important both for the advance speed of the lobe through the ICM and for determining the shape of the lobe. The fact that the lobe is near pressure balance with its environment means that the ICM can influence the shape of the lobe. The relatively high density in the inner regions of the cluster will compress the lobe near the nucleus, separating it into the familiar double system and supporting model C in \citet{scu74}. Lobes near pressure balance can also be moulded by variations in the ICM, giving disturbances and asymmetries that are again familiar in observed systems. Given these observational results, the departure of our lobe length/volume ratio from that predicted by self-similar theory \citep{kai97} is to be expected. A useful consequence of the lobes being near pressure balance is that the lobe internal pressure can be estimated from the external pressure at the lobe mid-point if IC flux is not available for a more detailed calculation. The relationship between lobe tip Mach number and environment richness gives evidence of the lobe expansion being affected by the ICM, which could suggest that the amount of energy being inserted into the ICM for a given jet power is also affected by its richness. However, the simulations of \citet{hak13} suggest that although the Mach number is lower in a rich environment, the shock front spreads further round the lobe tip and so is physically larger. Thus they find that for jets of the same power, the amount of energy being input into the ICM is near independent of the richness of the environment. Our results will provide a useful benchmark for more detailed investigations of how the energy input is distributed within the ICM for different populations of radio galaxies.

One reason this is important is because the space density of high excitation radio galaxies (predominantly FRIIs) increases with redshift \citep{bes14}, with an increasing radio-loud fraction in lower mass galaxies at earlier times \citep{wil15}. The cumulative energy input into groups and clusters from such episodes is not well constrained, but may play a role in explaining the observed properties of the ICM: it is well established that scaling relations for galaxy clusters do not follow the self-similar relation predicted for purely gravitational heating (e.g. \citealt{edg91,mar98,pra09}). This indicates a need for processes such as AGN heating to break the similarity by affecting low-mass environments more strongly; however, the locations, mechanisms and evolution of this energy input is uncertain (e.g. \citealt{sho10,fab12,pik14}). Radio surveys such as LOFAR will soon make it possible to study more typical radio galaxies at the epoch of group and cluster formation, but inferring the impact of those populations on their environments will require well-calibrated dynamical models or scaling relations that take into account the effects of environment. Our combined measurements of FRII internal conditions and environment provide important constraints for these models.

The facts that we found a strong correlation between jet power and radio luminosity and that our relationship reflects the theoretical relationship of \citet{wil99} are encouraging. We estimated jet power using a new, independent method based on the internal lobe conditions and the lobe advance speed; the correlation was strong even when the common distance dependence was taken into account (see \citealt{god16}). This method is applicable to FRII galaxies in their later stages, where the particle content can be assumed to be dominated by relativistic leptons and the lobe tip is overpressured with respect to its environment, advancing with supersonic or near-supersonic speeds. It could provide a better estimate of the jet power/radio luminosity relationship for this population of radio galaxies than is currently available, which may be of use in estimating the impact of the high redshift HERG population on the environment. 

\subsection{Systematic uncertainties}\label{sec:errors}

We made a number of assumptions and approximations during the calculations which could impact on our findings. The following section looks at the likely size of the effects.

\subsubsection{Lobe volumes}\label{sec:lobeV}

Since the electron density scales with lobe volume ($V^{-4/7}$, \citealt{har04a} -- assuming an electron energy index of 2), a potentially large source of error is the definition of the lobe shapes and the consequent flux measurements and volume calculations. We used the same shapes and exclusions for both the radio and X-ray flux measurements, and subtracted any exclusions from the total volume. However, the shapes are of necessity simple, 2-dimensional shapes approximating a complex, 3-dimensional structure, which gives scope for large inaccuracies.

We looked at the difference in using a cylindrical lobe instead of ellipsoidal for 3C~452; this increased the volume by 50~per~cent and reduced the observed pressure by 30~per~cent. Using an ellipsoidal lobe instead of cylindrical for 3C~35 reduced the volume by 40~per~cent and increased the observed pressure by 60~per~cent. An increase or decrease of 60~per~cent in the observed pressure would, for example, increase/decrease a Mach number of 5 to 6.3/3.2 or a Mach number of 1 to 1.2/0.7. Because the size of the change decreases with Mach number this is unlikely to make a significant difference to the results. Note that both these tests gave a visibly poor fit to the radio lobes so are likely to be over-estimating the true uncertainty.

We also looked at the effect of inaccuracies in exclusion sizes by doubling and halving the exclusion volume estimates for four sources. The exclusion volumes ranged from 5~per~cent to 15~per~cent of the total lobe volume and gave similar percentage changes in the observed pressure. A 15~per~cent increase/decrease at Mach 5 would give a Mach number of 5.4/4.6, with the size of the change reducing with Mach number.

\subsubsection{Low-frequency spectral index}\label{sec:delta}

In previous inverse Compton studies (e.g. \citealt{har00,har02b,cro05b}), the electron energy distribution at low frequencies has been assumed to be a power law with an index of $\delta\sim2$, steepening to $\delta\sim2.5$ above a break frequency and then dropping exponentially at the cut-off frequency, following the theoretical work of \citet{hea87}. However, recent observations have allowed the low energy electron index to be measured down to $\sim30$~MHz for some sources (e.g. \citealt{harw13,harw15,harw16}), giving a steeper index more in line with the $178-750$~MHz spectral indices for the 3CRR sample\footnote{http://3crr.extragalactic.info}. We therefore used an electron energy index $\delta=2.4$ (spectral index $\alpha=0.7$) to calculate the equipartition and observed lobe fields and energy densities for all sources.

If the index is too steep, the energy density (and consequently the internal lobe pressure and Mach number) will be overestimated. We looked at the effect of a lower spectral index on the mid-lobe pressure ratio and Mach number for seven of the 3CRR sources with ratios of observed to external pressure ranging between 0.5 and 12. We found that reducing $\delta$ to 2.0 reduced the pressure ratio by 25~per~cent to 30~per~cent. A decrease of 30~per~cent would change Mach 5 to Mach 4.2 and Mach 1 to 0.9. As with the volume changes, if we have over-estimated the electron index it is unlikely to have much effect on the results.

If the index is too shallow, then the energy density will be underestimated. \citet{harw16} found the spectral index for one of our sources, 3C~452, to be 0.85 ($\delta=2.7$) and calculated the energy density for the combined lobes to be $\sim3\times10^{-12}$~J~m$^{-3}$ -- about three times higher than the energy densities we obtained using $\delta=2.4$. We recalculated our energy density for 3C~452 using $\delta=2.7$ (although not changing the other radio data in our calculations to match those used by \citealt{harw16}), and this gave an energy density of $2\times10^{-12}$~J~m$^{-3}$, much closer to the value found by Harwood~et~al.

We therefore estimated the low frequency spectral index $\alpha$ for 5 of our sources which did not have strong radio emission from the nucleus, jets and hot-spots. For this we used published low frequency flux densities (at 178~MHz\footnote{http://3crr.extragalactic.info} or 408~MHz\footnote{http://vizier.cfa.harvard.edu/viz-bin/VizieR?-source=VIII/16}) together with the flux densities obtained from the radio maps used in this study (ranging between 0.6 to 4.8 GHz). $\alpha$ ranged from 0.77 to 0.94, well within the range of low frequency spectral indices published for the 3CRR sample. Our spectral indices give $\delta$ from 2.5 to 2.9, all higher than the value of 2.4 used in our analysis. A steeper index gives a higher internal lobe pressure and consequently a higher Mach number -- for the source with the highest electron index (3C~472.1), the Mach number increased from 1.1 to 2.3.

The choice of electron energy index could therefore have a fairly large effect on the calculation of lobe internal energy density and pressure, and consequently on the Mach number. However, if all the indices are higher than 2.4, as with our test cases, then all internal pressures and Mach numbers will be higher than estimated here and so the overall study results will remain the same. Since our Mach numbers cover a similar range to those in the studies cited in Section~\ref{sec:intro}, it seems unlikely that the indices sufficiently far in error to make a big difference to our results.

\subsubsection{Maximum and minimum energies}\label{sec:gamma}

We followed \citet{cro05b} in using Lorentz factors of $\gamma_{\rm min}=10$ and $\gamma_{\rm max}=10^5$ to define the minimum and maximum electron energies for the inverse Compton calculations. As discussed by \citet{cro05b}, it is the low energy population of electrons that scatter the CMB photons (which are responsible for the bulk of the lobe X-ray emission) to the observed X-ray frequencies, so varying $\gamma_{\rm max}$ makes little difference. They chose the conservative value of $\gamma_{\rm min}=10$, an order of magnitide lower than had previously been observed in hot-spots, but found that varying it between 10 and 1000 gave the same results to within the $1\sigma$ errors.

\subsubsection{Radio observation properties}

There was no consistent set of low-frequency radio maps for the sources in the samples. Since the commonest map frequency was 1.4~GHz, when possible we used these to obtain the flux density. For sources with maps at a higher frequency than 1.4~GHz, we used published flux density measurements at lower frequencies as listed in Table~\ref{tab:flux}, since the lower frequencies were less likely to be contaminated by nuclear and hot-spot emission.

There were also differences in the resolutions of the radio maps, which could have an effect on the definition of the lobe shapes. In particular, high resolution maps are likely to have a small LAS and miss regions of diffuse gas.

Two-thirds of the radio observations fell into two groups -- those with 1.4 GHz observations and resolutions between 1-5 arcsec (21 lobes) and those with 5 or 8.5 GHz observations again with resolutions between 1-5 arcsec (10 lobes). These two groups had almost identical ranges of source size (40 to 300 kpc) and similar medians (150 and 167 kpc respectively). We then looked at the ratio of observed to equipartition magnetic fields ($B_{\rm obs}/B_{\rm eqp}$). The 1.4 GHz subsample contained both the maximum and minimum $B_{\rm obs}/B_{\rm eqp}$ from the full sample (median 0.40, range 0.07-0.83). The high frequency subsample had a median $B_{\rm obs}/B_{\rm eqp}$ of 0.47 and a range of 0.24 to 0.67. A Kolmogorov-Smirnov test showed that there was no significant difference between the two distributions ($D=0.39$, $p=0.16$). Thus for the bulk of our sources the difference in frequency does not seem to be a problem.

Of the remaining lobes, five had 1.4 GHz observations at resolutions greater than 10 arcsec. These were large sources with lobe lengths from 290 to 1330~kpc. Their $B_{\rm obs}/B_{\rm eqp}$ ratios were however similar to those of the two higher resolution samples (median 0.39, range 0.22 to 0.65). All but one of the remaining sources had frequencies and resolutions lying between those of the three groups already described. Their lobe lengths range between 67 and 310~kpc and their $B_{\rm obs}/B_{\rm eqp}$ ratios have a median of 0.40, range 0.25 to 0.57. Again, there is nothing to distinguish their results from those of the other sources. The final source has low map frequency (610 MHz) and resolution (30 arcsec) and is large (550~kpc) but the $B_{\rm obs}/B_{\rm eqp}$ ratio is near the median (0.41).

Overall, it seems unlikely that the range of frequencies and resolutions used in this study had much effect on the results.

\subsubsection{Lobe viewing angle}

We do not know the viewing angle of the radio lobe. In standard unified models \citep{bar89,har98c} Narrow Line Radio Galaxies (NLRGs) have viewing angles of greater than $\sim$45$^\circ$; Broad Line Radio Galaxies (BLRGs) and Quasars (QSOs) have viewing angle of less than $\sim45^\circ$. A small viewing angle will decrease the apparent volume of the lobe, increasing the calculated energy density and the observed lobe pressure. However, reducing the viewing angle also reduces the angular distance to mid-lobe and the lobe tip, so the ICM density will also be overestimated. 

We looked at the effect of viewing angle on NLRG 3C~98 (N lobe) and BLRG PKS~0945+07 (W lobe). For 3C~98, we assumed a viewing angle of 45$^\circ$ and recalculated the volume, ICM pressures and lobe observed and equipartition pressures. This brought the lobe nearer equipartition ($P_{\rm obs}/P_{\rm eqp}$ reduced from 3.1 to 2.6). The pressure balance was increased (by 16~per~cent at mid-lobe and 13~per~cent at the lobe tip). This size of change had only a slight effect on the Mach number, raising it from 4.1 to 4.4.

For PKS~0945+07 we assumed a viewing angle of 20$^\circ$. Again, the effect of increasing the volume estimate brought the lobe nearer to equipartition ($P_{\rm obs}/P_{\rm eqp}$ reduced from 5.7 to 3.2). In this case the pressure balance was almost unchanged, with the Mach number changing by less than 1~per~cent.

The relative changes in lobe and ICM pressures will depend on the environment richness and shape -- both 3C~98 and PKS~0945+07 have $\beta$-model parameters below the median so the ICM pressure will drop more slowly than for a steep $\beta$-model source. We therefore looked also at 3C~285, an NLRG with a $\beta$-model that is steeper than the median. The changes were smaller than for the shallow model, with only a 4~per~cent increase in pressure balance at the lobe tip. We therefore agree with the conclusions of \citet{har00} and \citet{cro04} that the viewing angle has a similar effect on the calculated pressures in the lobe and the ICM. It will therefore have only a small effect on the calculation of the pressure ratios and Mach number.

\subsubsection{Excluded lobes}\label{sec:excl}

As mentioned in Section~\ref{sec:sample}, results could not be calculated for a number of lobes. There were two main reasons for exclusion:

a) The lobes were so small in angular extent that they were masked by nuclear and/or central ICM emission. These lobes were either very young, in which case they may still be evolving rapidly and not be representative of the population of stable lobes (e.g. \citealt{hak13}), or they are at a shallow angle to the observer. In the latter case, the sample is large enough that it is unlikely that the lobes are different from the rest of the sample. A large lobe, 1~Mpc for example, at a viewing angle of 20$^\circ$ would still appear more than 300~kpc long. The largest of our excluded lobes was 75~kpc; even at a pessimistic viewing angle of 20$^\circ$ the lobes are still only $\sim$200~kpc long and so we were not excluding scarce large lobes.

b) The lobes were only partially on the observing chip, or, in one case, were over a 4-chip join. These lobes are likely to be amongst the largest in angular extent, and potentially in physical extent and so might bias the sample. Of the five sources with off-chip lobes, only one had both lobes off chip so all the others had their second lobe in the sample. The source with both lobes off-chip was the largest of this group of excluded sources with lobes of $\sim$1~Mpc, but it also lay in a very weak ICM with upper limits on the external pressure so we could not have calculated the Mach number for this source.

We also excluded 3C~236 because the lobes were so large (2.7 and 1.9~Mpc) that we did not trust the extrapolation of the ICM $\beta$ model to provide the pressures. This is by far the biggest source in the sample -- the next largest lobe is 1.3 ~Mpc long. In all, only three of the 39 lobes for which we could calculate Mach numbers were longer than 0.5~Mpc, and only one lobe was longer than 1~Mpc. There is therefore the possibility of bias against large lobes; having said that, we have Mach numbers for lobes from 40 to 1300~kpc long, which is a wide range of lobe sizes, and the few large lobes in the sample do not stand out as being different from the rest of the sample.

\subsubsection{Environment measurements}

A potential problem with the measurements of the external environment comes from the extrapolation of the $\beta$ model beyond $D_{\rm rad}$ for the ICM pressure calculations. There were 10 lobes where the lobe tip position was further from the cluster centre than $D_{\rm rad}$. These all have environments with poorly constrained $\beta$ models and so their external pressures have large errors. 3C~321 has the largest lobe-tip distances in the sample and was discussed in Section~\ref{sec:M2} -- its lobes have unreasonably large Mach numbers and all statistics were calculated with and without these two lobes. 3C~326's lobes have the lowest Mach numbers (0.7) -- again these are large lobes so extrapolation from an inaccurate $\beta$ model could be overestimating the ICM pressure. 3C~33 has relatively small lobes but is in a weak environment. Its Mach numbers are quite high (4 and 3.5) but reasonable for such a weak environment and there are two others higher. The other 4 lobes all have Mach numbers near the median and so do not stand out in any way.

We are again a little concerned at possible errors involving large lobes -- the highest and lowest Mach numbers come from the largest sources in the sample -- but do not think that the overall results will be much affected.

\subsubsection{ICM temperature variation}

As discussed in \citet{ine13,ine15}, the ICM temperatures used in the analysis were an average temperature taken across a wide radius. Thus they are representative of the overall temperature, and not the temperature at a specific point in the profile. We had very few sources with enough counts to generate a temperature profile. Typically, temperatures rise steeply from the cluster centre to a peak about 20~per~cent higher than the average temperature (e.g. \citealt{ras07}). This occurs at a radius of $0.1R_{500}$ for groups, rising to $0.15R_{500}$ for groups and clusters with average temperatures of $\sim2$~keV and above. The temperatures then fall gradually, reaching about two-thirds of the average temperature by $R_{500}$. We may therefore be over-estimating the external pressure at the lobe tips, and consequently under-estimating the Mach number.

This sample of radio lobes cover a wide range of sizes. The median radii of the mid-points and tips are at about $0.2R_{500}$ and $0.4R_{500}$ respectively. At these radii the temperature is very similar to the average. Only one lobe has its mid-point beyond $R_{500}$ -- this is 3C~326, and the $\beta$-model extrapolation for that radius gives a 1$\sigma$ lower bound close to zero. There are 7 lobe tips from our sample of 47 that extend beyond $R_{500}$ (including 3C~326). Of these, 5 have such large errors from the $\beta$-model extrapolation that any pressure variation due to temperature is small in comparison; the other two have 1$\sigma$ errors of a similar size to the expected change in pressure from the temperature variation. We may therefore be under-estimating the Mach number for these sources, but it is unlikely that the temperature variation with radius will have any effect on our overall results.

\subsubsection{Jet ram pressures}\label{sec:ram}

We neglected the jet ram pressure when calculating the Mach number, and so it is useful to ask whether the Mach numbers are underestimated. The momentum flux up the jet always provides an additional pressure term, but the standard assumption is that it is distributed over the whole leading edge of the lobe, both by local hydrodynamic effects and by the fact that the jet termination probably moves about in the lobe as discussed in Section~\ref{sec:M2}. The ram pressure exerted by the jet on the end of the lobes is therefore lower than the ram pressure on a test surface {\it within} the jet by a large factor -- the ratio of the lobe cross-sectional area to the jet cross-sectional area.

We estimated ram pressures for some of the lobes from the jet power $Q_{\rm jet}$, jet velocity $c$ (the speed of light, assuming a light, relativistic jet) and surface area $A$ of the lobe tip, using $P_{\rm ram}=Q_{\rm jet}/(cA)$. This is extremely sensitive to the area estimate, so we only looked at lobes where the map quality and resolution were good and the lobe shape was not complex -- this left 11 lobes for which we could estimate the area of the lobe tip (the Mach numbers ranged from 0.7 to 4.8). For cylindrical lobes, we used the flat area of the cylinder end, and for elliptical lobes we estimated the surface area of the forward end of the ellipse pushing into the ICM. The highest ratio of ram pressure to internal lobe pressures was 0.06, suggesting that including ram pressure in the Mach number calculations would have little effect.

\subsubsection{Overall effect of systematics}

Overall, the estimates of lobe volumes and the assumption of a constant low-frequency spectral index are likely to be the biggest sources of error in the results. Improved lobe volume estimates will need better resolution low-frequency radio maps, but it would be possible to make estimates of spectral indices where data are available, look at the overall variation and make a better assessment of the impact of the assumption of a constant index of 0.7. However, the strength of the main results is such that it is unlikely that any of the systematic errors described here will modify them significantly.

\section{Conclusions}

We have carried out the first comprehensive X-ray study of a large representative sample of the lobes of FRII radio galaxies, at redshifts 0.1 and 0.5. We used measurements of X-ray lobe inverse-Compton emission to determine the internal pressures within the radio lobes, and compared them with the equipartition (minimum energy) values, and we determined the external pressures acting on the radio lobes from the thermal X-ray emission from the ICM, which was characterized for this sample in \citet{ine15}. These results were used to investigate the lobe dynamics, estimating the Mach numbers for lobe expansion, and estimating the jet power for each source based on the measured internal energy and inferred expansion speed.

Our main conclusions are:
\begin{itemize}
\item All lobe internal energies are higher than those predicted by equipartition (in the absence of protons), but all are within one order of magnitude of the equipartition prediction. This is consistent with previous studies, and demonstrates a clear difference in how the internal energy is distributed between particles and field compared to the FRI population, which requires a substantial proton population.

\item Almost all lobes were calculated to be over-pressured at the tip compared with the ICM, as expected for expanding lobes;

\item Lobe pressures at mid-lobe were near pressure balance with the ICM (within one order of magnitude of the ICM pressure for all but three lobes), allowing the ICM to shape the regions of the lobes near the nucleus as they expand outwards;

\item Lobe tip Mach numbers were below 5 for all but one source, with a median of 1.8. Given that the Mach numbers are lower limits, it is likely that at least half the sample are driving strong shocks into the ICM;

\item We found a jet power -- radio-luminosity relation whose slope is in good agreement with the relation of \citet{wil99}, with a higher normalisation, broadly as expected given the departure from equipartition and distribution of environments we find for our sample.
\end{itemize}

Overall, these results provide a useful step towards characterising relationships between the properties of FRII radio galaxies and their environments and will be helpful for developing theories of galaxy and cluster evolution and for calibrating simulations. 

\clearpage
\input{tab_rawdata}
\input{tab_excl}

\input{tab_obsdata}
\input{tab_beta}
\input{tab_lobes}
\input{tab_flux}
\input{tab_BandP}
\input{tab_KTau}

\section*{Acknowledgements}
The authors thank the anonymous referee for the kind and useful comments on this paper. JI acknowledges the support of the South-East Physics Network (SEPnet). JHC acknowledges support from the UK Science and Technology Facilities Council (STFC) under grant ST/M001326/1. MJH acknowledges support from the STFC under grant ST/M001008/1.

The scientific results reported in this article are based on observations made with the {\it Chandra} X-ray observatory and on observations obtained with {\it XMM-Newton}, an ESA science mission with instruments and contributions directly funded by ESA Member States and NASA. This research has made use of software provided by the {\it Chandra} X-ray Center (CXC) in the application package \textsc{ciao}, and of the {\it XMM-Newton} Science Analysis Software (\textsc{sas}). Figures were plotted using the \textit{veusz} package (http://home.gna.org/veusz/).




\bibliographystyle{mnras}
\bibliography{Bibliography}






\bsp	
\label{lastpage}
\end{document}

%% file: tab_rawdata.tex
\begin{table*}
\caption{The sample of FRII sources.}
\begin{tabular}{lllccccc}\hline
Source&\multicolumn{2}{c}{RA  (J2000)   Dec}&Redshift&Scale&log$_{10}L_{151}$&Type&$N_{\rm H}$\\
& h m s &deg min sec&&kpc arcsec$^{-1}$&W~Hz$^{-1}$~sr$^{-1}$&&x10$^{20}$~cm$^{-2}$\\
\hline
3C 33&01 08 52.86&+13 20 14.2&0.060&1.15&25.93&HERG&3.90\\
3C 35&01 12 02.26&+49 28 35.5&0.067&1.28&25.34&LERG&13.00\\
3C 98&03 58 54.43&+10 26 02.8&0.031&0.62&25.28&HERG&15.10\\
3C 192&08 05 35.01&+24 09 49.7&0.060&1.15&25.53&HERG&4.21\\
3C 219&09 21 08.63&+45 38 57.3&0.174&2.95&26.52&HERG&1.51\\
4C 73.08&09 49 45.78&+73 14 23.1&0.059&1.14&25.34&HERG&2.29\\
3C 285&13 21 17.86&+42 35 14.8&0.079&1.50&25.51&HERG&1.27\\
3C 303&14 43 02.76&+52 01 37.2&0.141&2.48&25.75&HERG&1.58\\
3C 321&15 31 43.46&+24 04 19.0&0.096&1.78&25.76&HERG&4.11\\
3C 326&15 52 09.10&+20 05 48.3&0.089&1.67&25.87&LERG&3.81\\
3C 433&21 23 44.56&+25 04 28.0&0.102&1.88&26.15&HERG&11.90\\
3C 452&22 45 48.75&+39 41 15.9&0.081&1.53&26.21&HERG&11.30\\

PKS 0034$-$01&00 37 49.18&-01 09 08.2&0.073&1.40&25.54&LERG&3.07\\
PKS 0038+09&00 40 50.53&+10 03 26.8&0.188&3.14&26.44&HERG&5.51\\
PKS 0043$-$42&00 46 17.75&-42 07 51.4&0.116&2.10&26.23&LERG&2.21\\
PKS 0213$-$13&02 15 37.5&-12 59 30.5&0.147&2.57&26.23&HERG&1.92\\
PKS 0349$-$27&03 51 35.81&-27 44 33.8&0.066&1.26&25.55&HERG&0.99\\
PKS 0404+03&04 07 16.49&+03 42 25.8&0.089&1.66&25.89&HERG&11.90\\
PKS 0806$-$10&08 08 53.600&-10 27 39.71&0.109&1.99&25.92&HERG&7.74\\
PKS 0945+07&09 47 45.15&+07 25 20.4&0.086&1.62&25.91&HERG&3.00\\
PKS 1559+02&16 02 27.38&+01 57 55.7&0.104&1.91&26.12&HERG&6.44\\
PKS 2221$-$02&22 23 49.57&-02 08 12.4&0.056&1.09&25.48&HERG&4.87\\
PKS 2356$-$61&23 59 04.50&-60 54 59.1&0.096&1.78&26.27&HERG&2.38\\
\hline
3C 16&00 37 45.39&+13 20 09.6&0.405&5.41&26.82&HERG&4.48\\
3C 46&01 35 28.47&+37 54 05.7&0.437&5.66&26.84&HERG&5.66\\
3C 200&08 27 25.38&+29 18 45.5&0.458&5.82&26.92&LERG&3.74\\
3C 228&09 50 10.79&+14 20 00.9&0.552&6.42&27.37&HERG&3.18\\
3C 244.1&10 33 33.97&+58 14 35.8&0.430&5.61&27.10&HERG&0.58\\
3C 274.1&12 35 26.64&+21 20 34.7&0.422&5.55&27.02&HERG&2.00\\
3C 330&16 09 35.01&+65 56 37.7&0.549&6.41&27.43&HERG&2.81\\
3C 427.1&21 04 07.07&+76 33 10.8&0.572&6.54&27.53&LERG&10.90\\
3C 457&23 12 07.57&+18 45 41.4&0.428&5.59&27.00&HERG&22.3\\
6C 0850+3747&08 50 24.77&+37 47 09.1&0.407&5.43&26.15&HERG&2.95\\
6C 0857+3945&08 57 43.56&+39 45 29.0&0.528&6.28&26.34&HERG&2.64\\
6C 1132+3439&11 32 45.74&+34 39 36.2&0.512&6.18&26.33&HERG&2.14\\
6C 1200+3416&12 00 53.34&+34 16 47.3&0.530&6.29&26.17&LERG&1.62\\
7C 0219+3423&02 19 37.83&+34 23 11.2&0.595&6.66&25.98&HERG?&6.30\\
\hline\end{tabular}

Column 1: source name; Cols. 2-3: right ascension and declination, J2000 coordinates; Col. 4: redshift; Col. 5: angular scale; Col. 6: 151 MHz radio luminosity; Col. 7: radio-loud AGN spectral class; Col. 8: AGN morphology; Col. 9: Column density.
\label{tab:rawdata}\end{table*}

%% file: tab_excl.tex
\begin{table*}
\caption{FRII lobes in the original samples that were excluded from this study}
\begin{footnotesize}\begin{minipage}[t]{1.0\textwidth}

\begin{tabular}{llp{10cm}}\hline\hline
Source&Lobe&Reason for exclusion\\
\hline
3C~28&Both&Small lobes in a rich, disturbed environment\\
3C~98&South&Lobe is across a 4-chip join\\
DA~240&Both&Both lobes partially off the chip\\
4C~73.08&East&Lobe is partially off the chip\\
3C~236&Both&Large lobes (2.7 and 1.9~Mpc) which extend substantially beyond both the maximum detected radius (0.3~Mpc) and the $R_{200}$ overdensity radius (0.7~Mpc)\\
3C~388&Both&Lobes are in a strong, disturbed environment\\
3C~390.3&Both&Lobes are in a strong, disturbed environment\\
3C~433&North&North lobe has FRI morphology \citep{van83,hod10}\\
PKS~0349$-$27&North&Lobe is across chip boundary\\
PKS~0442$-$28&Both&Poor quality radio map\\
PKS~0945+07&East&Lobe lies across read-out streak and chip boundary\\
PKS~1559+02&West&Lobe is partially off the chip\\
PKS~1733$-$56&Both&Poor quality radio map\\
PKS~1949+02&Both&Complex, X-shaped lobes with no apparent IC emission\\
PKS~2211$-$17&Both&Lobes are in a strong, disturbed environment, but used Mach number from \citet{cro11} in Mach number analysis\\
PKS~2221$-$17&North&Lobe is partially off the chip\\
\hline
3C~19&Both&Small lobes (4 arcsec) in a strong environment\\
3C~295&Both&Small lobes (3.5 arcsec) in a strong environment\\
6C~0850+3747&North&Lobe is over the nucleus\\
7C~0213&Both&Low resolution map which was not good enough to define the lobe shape\\
7C~0223&Both&Low resolution map which was not good enough to define the lobe shape\\
7C~1731&Both&Lobes are less than 1 arcsec in extent\\
TOOT~1303+3334&Both&No lobe structure visible\\

\hline\end{tabular}
\end{minipage}\end{footnotesize}\label{tab:excl}\end{table*}

%% file: tab_obsdata.tex
\begin{table*}
\caption{Observational data for the sample.}
\begin{tabular}{lccccccl}\hline
Source&X-ray${^a}$&Observation&Exposure${^b}$&Screened${^b}$&Radio map&Resolution&Ref.\\
&Instrument&ID&time (ks)&time (ks)&freq. (GHz)&(arcsec)\\
\hline
3C 33&C&6910, 7200&39.83&39.61&1.5&$4 \times 4$&1\\
3C 35&C&10240&25.63&25.63&1.4&$14 \times 12$&1\\
3C 98&C&10234&31.71&31.71&4.9&$3.7 \times 3.7$&1\\
3C 192&C&9270&10.02&9.62&1.4&$3.9 \times 3.9$&1\\
3C 219&C&827&19.24&16.79&1.5&$1.4 \times 1.4$&1\\
4C 73.08&C&10239&28.52&28.52&0.61&$30 \times 30$&1\\
3C 285&C&6911&39.62&39.61&1.5&$5.5 \times 5.5$&1\\
3C 303&C&1623&15.10&14.95&1.5&$1.2 \times 1.2$&1\\
3C 321&C&3138&47.13&46.87&1.5&$1.4 \times 1.4$&1\\
3C 326&C&10908, 10242&45.81&45.81&1.4&$14 \times 39$&1\\
3C 433&C&7881&38.19&37.15&8.5&$0.75 \times 0.75$&1\\
3C 452&C&2195&79.92&79.53&1.4&$6 \times 6$&1\\
PKS 0034$-$01&C&2178&27.52&26.54&4.9&$4.5 \times 3.7$&2\\
PKS 0038+09&C&9293&7.94&7.94&4.9&$4.4 \times 3.4$&2\\
PKS 0043$-$42&C&10319&18.38&18.38&8.6&$1.2 \times 0.88$&3\\
PKS 0213$-$13&C&10320&19.89&19.89&4.9&$5.9 \times 3.4$&2\\
PKS 0349$-$27&C&11497&19.89&19.89&1.5&$11 \times 8.9$&4\\
PKS 0404+03&C&9299&8.07&8.07&8.4&$2.2 \times 2.2$&5\\
PKS 0806$-$10&C&11501&19.79&19.79&4.9&$6.8 \times 1.6$&4\\
PKS 0915$-$11&C&4970&98.82&98.42&1.4&$2 \times 1.5$&4\\
PKS 0945+07&C&6842&29.78&29.78&1.5&$4 \times 4$&6\\
PKS 1559+02&C&6841&39.65&39.62&8.5&$2.2 \times 2.2$&5\\
PKS 2221$-$02&C&7869&45.60&45.60&8.2&$2.4 \times 2.4$&5\\
PKS 2356$-$61&C&11507&19.79&19.79&1.5&$7.2 \times 6.9$&7\\
\hline
3C 16&C&13879&11.9&11.9&1.4&$1.2 \times 1.2$&1\\
3C 46&X&0600450501&17.9&5.5&1.5&$4.2 \times 4.2$&1\\
3C 200&C&838&14.7&14.7&4.9&$0.33 \times 0.33$&1\\
3C 228&C&2453/2095&10.6/13.8&24.4&8.4&$1.2 \times 1.2$&8\\
3C 244.1&C&13882&11.9&11.8&8.4&$0.75 \times 0.75$&8\\
3C 274.1&X&0671640801&27.2&22.5&$1.4$&$5.4 \times 5.4$&9\\
3C 330&C&2127&44.2&44.0&1.5&$1.5 \times 1.5$&10\\
3C 427.1&C&2194&39.5&39.5&1.5&$1.8 \times 1.1$&11\\
3C 457&X&0502500101&52.2&36.8&1.4&$5.1 \times 5.1$&1\\
6C 0850+3747&C&11576&39.2&39.2&$1.4$&$1.4 \times 1.3$&4\\
6C 0857+3945&X&551630601&24.9&10.0&$1.4$&$5.4 \times 5.4$&9\\
6C 1132+3439&C&11577&39.6&39.6&$1.4$&$5.4 \times 5.4$&9\\
6C 1200+3416&X&0551630301&49.6&37.6&$1.4$&$5.4 \times 5.4$&9\\
7C 0219+3423&C&11575&39.3&39.3&$1.4$&$1.4 \times 1.3$&4\\
\hline\end{tabular}

References: (1) http://www.jb.man.ac.uk/atlas, (2) \citet{mor93}, (3) \citet{mor99}, (4) Made from the VLA archives \citep{mit05}, (5) \citet{lea97}, (6) \citet{har07a}, (7) Made from the ATCA archives, (8) \citet{mul08}, (9) \citet{bec95}, (10) \citet{har02b}, (11) \citet{cro05b}\\

${^a}$ C=\textit{Chandra}, X=\textit{XMM-Newton}.
${^b}$ pn camera times for {\it XMM-Newton} sources.
\label{tab:obsdata}\end{table*}

%% file: tab_beta.tex
\begin{table*}
\caption{Radial profile modelling for the ICM.}
\begin{tabular}{lcccllcll}\hline
Source&$D_{\rm rad}$$^a$&Counts$^{ b}$&\multicolumn{3}{l}{ICM (outer) model$^{ c,d}$}&\multicolumn{3}{l}{Host galaxy (inner) model$^{ c}$}\\
&kpc&&$\chi^2$/dof&$\beta$&$r_{\rm c}$ (kpc)&$\chi^2$/dof&$\beta$&$r_{\rm c}$ (kpc)\\
\hline
3C 33&113&2770&3.2/8&0.76 (0.30--1.20)&16 (112--1)&4.4/6&1.20 (0.85--2.00)&1.0 (2.0--0.5)\\
3C 35&221&218&3.5/5&1.17 (0.30--1.20)&134 (384--2)&&&\\
3C 98&121&1380&1.4/8&0.42 (0.30--1.20)&1.8 (58.1--1.0)&1.4/6&2.92 (0.77--3.00)&0.9 (1.3--0.2)\\
3C 192&170&191&1.1/3&0.41 (0.30--0.91)&1.0 (10.6--1.0)&&&\\
3C 219&726&2251&19/9&0.40 (0.31--0.59)&28 (90--4)&&&\\
4C 73.08&167&624&8.9/4&0.42 (0.31--1.20)&1.0 (68.3--1.0)&&&\\
3C 285&368&1521&4.1/9&0.36 (0.32--0.70)&14 (82--5)&4.3/7&1.12 (1.00--3.00)&0.8 (4.1--0.5)\\
3C 303&366&2510&3.7/9&0.51 (0.44--0.70)&1.2 (10.3--1.0)&&&\\
3C 321&87&843&6.8/8&1.19 (0.31--1.20)&30 (55--5)&7.2/6&2.64 (1.07--3.50)&6.1 (8.3--2.8)\\
3C 326&328&321&0.2/5&1.11 (0.30--1.20)&738 (1670--10)&0.2/3&0.95 (0.70--1.50)&10 (35--4)\\
3C 433&323&3058&12.2/10&1.09 (0.30--1.20)&310 (593--3)&12.3/8&1.14 (0.50--1.20)&10 (16--1)\\
3C 452&300&3202&7.4/12&0.74 (0.42--1.20)&64 (125--21)&7.5/10&1.52 (0.96--3.00)&1.3 (2.5--0.6)\\
PKS 0034$-$01&&$<$254&&{\it 0.47}&{\it 40.90}&&&\\
PKS 0038+09&231&1238&43/9&0.78 (0.30--1.20)&71 (975--3)&43/7&2.50 (0.96--2.50)&19 (25--8)\\
PKS 0043$-$42&413&576&7.8/5&0.34 (0.30--0.44)&1.0 (20.7--1.0)&&&\\
PKS 0213$-$13&126&1244&2.9/5&0.63 (0.30--1.20)&18 (81--4)&&&\\
PKS 0349$-$27&310&839&1.0/8&0.30 (0.30--1.50)&97 (1256--39)&1.1/6&2.13 (0.45--2.50)&11 (23--1)\\
PKS 0404+03&&$<$122&&{\it 0.47}&{\it 40.89}&&&\\
PKS 0806$-$10&&$<$161&&{\it 0.47}&{\it 40.99}&&&\\
PKS 0945+07&318&4404&10/13&0.40 (0.31--0.95)&7.0 (50.1--2.0)&11/11&0.86 (0.71--1.40)&0.6 (1.5--0.4)\\
PKS 1559+02&564&1255&8.7/11&0.30 (0.30--1.20)&6.9 (1807.8--1.0)&9.0/9&2.99 (1.14--3.00)&3.3 (3.8--1.3)\\
PKS 2221$-$02&240&8775&22/16&0.35 (0.30--1.00)&16 (105--7)&28/14&1.15 (0.99--1.34)&1.5 (1.8--1.2)\\
PKS 2356$-$61&526&1791&1.8/14&0.42 (0.30--1.20)&56 (279--12)&1.9/12&1.16 (0.64--2.00)&3.9 (9.4--1.1)\\
\hline
3C 16&${\it (<60)}$&\multicolumn{2}{c}{{\it Low counts}}&{\it 0.49}&{\it 5.74} \\
3C 46&510&170&0.57/2&1.49 (0.52--1.50)&375 (477--118) \\
3C 200&200&259&0.77/3&0.48 (0.43--0.63)&6.4 (16.7--5.9) \\
3C 228&537&768&3.3/6&0.90 (0.57--2.0)&79.1 (179--39) \\
3C 244.1&966&171&0.82/3&0.41 (0.36--0.45)&5.78 (17.73--5.72) \\
3C 274.1&610&678&3.8/5&1.16 (0.75--1.50)&177 (241--103) \\
3C 330&473&360&1.3/7&0.56 (0.49--0.67)&30.8 (47.7--18.9) \\
3C 427.1&675&721&4.7/7&0.40 (0.38--0.54)&20.5 (28.5--12.8) \\
3C 457&1119&2402&6.5/6&0.45 (0.34--0.66)&102 (268--6) \\
6C 0850+3747&534&2351&10/6&0.45 (0.43--0.49)&5.48 (13.57--5.43) \\
6C 0857+3747&816&612&2.2/6&0.41 (0.24--0.56)&7.03 (106.3--6.3) \\
6C 1132+3439&669&389&9.3/7&0.38 (0.29--0.45)&47.8 (161.8--6.2) \\
6C 1200+3416&1007&1983&36/7&0.41 (0.37--0.44)&39.1 (39.6--9.9) \\
7C 0219+3423&164&46&0.008/1&0.45 (0.21--0.99)&31.3 (66.5--6.7) \\
\hline\end{tabular}

Column 1: source name. Col. 2: maximum detected radius. Col. 3: net counts in the surface brightness profile. Col. 4: $\chi^2$ and degrees of freedom for the ICM model. Col. 5: $\beta$ for the ICM model. Col. 6: core radius for the ICM model. Col. 7: $\chi^2$ and degrees of freedom for the inner model (if used). Col. 8: $\beta$ for the inner model. Col. 9: core radius for the inner model.

$^{a}$ Lower limits indicate that the detected ICM emission extended beyond the chip.
$^{b}$ Counts for {\it XMM-Newton} sources are for the pn camera only. Upper limits were obtained within estimated R$_{500}$.
$^{c}$ Values for $\beta$ and core radius $r_{\rm c}$ are best fit parameters. Ranges are the Bayesian credible intervals.
$^{d}$ Italics indicate median values used for sources with low counts.

\label{tab:beta}\end{table*}

%% file: tab_lobes.tex
\begin{table*}\centering
\caption{ICM properties and lobe sizes.}
\begin{footnotesize}\begin{minipage}[t]{1.0\textwidth}
\begin{tabular}{lcccccccc}\hline\hline
Source&ICM&ICM$^{a}$&Lobe$^{b}$&Lobe tip&Lobe-tip&Mid-lobe&Mid-lobe&Volume$^{c}$\\
&temperature&luminosity&&&ICM pressure&&ICM pressure&(radius)\\
&keV&$\times{10}^{43}$~erg~s$^{-1}$&&kpc&$\times{10}^{-14}$~Pa&kpc&$\times{10}^{-14}$~Pa&kpc\\
\hline
3C 33&$1.12^{+0.01}_{-0.02}$&$0.491^{+0.020}_{-0.027}$&N&150&$0.943^{+0.127}_{-0.942}$&86&$2.62^{+0.80}_{-2.34}$&60\\
3C 33&&&S&119&$1.43^{+0.26}_{-1.43}$&59&$5.27^{+2.18}_{-3.47}$&58\\
3C 35&$0.97^{+0.10}_{-0.20}$&$0.302^{+0.127}_{-0.174}$&N&468&$1.39^{+0.28}_{-1.26}$&234&$4.49^{+1.50}_{-2.07}$&185\\
3C 35&&&S&488&$1.29^{+0.26}_{-1.17}$&244&$4.17^{+1.40}_{-2.06}$&185\\
3C 98&$0.62^{+0.04}_{-0.06}$&$0.0660^{+0.0199}_{-0.0211}$&N&85&$2.47^{+0.91}_{-2.45}$&42&$7.26^{+3.89}_{-4.77}$&41\\
3C 192&$0.80^{+0.07}_{-0.12}$&$0.155^{+0.057}_{-0.067}$&C&126&$4.74^{+1.12}_{-1.03}$&63&$11.5^{+2.4}_{-1.7}$&64\\
3C 219&$1.46^{+0.13}_{-0.14}$&$5.19^{+0.58}_{-0.51}$&N&291&$24.4^{+1.1}_{-1.1}$&145&$54.2^{+2.2}_{-2.7}$&113\\
3C 219&&&S&280&$25.5^{+1.2}_{-1.2}$&145&$54.5^{+2.2}_{-2.7}$&107\\
4C 73.08&$1.37^{+0.26}_{-0.18}$&$0.0491^{+0.0292}_{-0.0418}$&W&550&$0.645^{+0.178}_{-0.645}$&275&$1.60^{+0.47}_{-1.60}$&237\\
3C 285&$0.94^{+0.10}_{-0.22}$&$0.559^{+0.135}_{-0.111}$&E&132&$10.1^{+0.6}_{-0.6}$&66&$22.1^{+1.4}_{-1.6}$&69\\
3C 285&&&W&172&$7.44^{+0.66}_{-0.52}$&86&$16.6^{+1.0}_{-1.1}$&84\\
3C 303&$0.94^{+0.09}_{-0.13}$&$0.757^{+0.115}_{-0.114}$&C&72&$24.4^{+3.1}_{-2.0}$&36&$72.3^{+7.0}_{-4.7}$&42\\
3C 321&$0.87^{+0.01}_{-0.02}$&$0.211^{+0.011}_{-0.013}$&N&272&$0.307^{+0.020}_{-0.281}$&246&$0.377^{+0.009}_{-0.338}$&17\\
3C 321&&&S&282&$0.284^{+0.023}_{-0.262}$&237&$0.387^{+0.009}_{-0.347}$&38\\
3C 326&$1.94^{+0.40}_{-1.35}$&$4.65^{+3.97}_{-4.59}$&E&741&$5.57^{+1.73}_{-5.56}$&370&$9.17^{+4.51}_{-3.12}$&302\\
3C 326&&&W&1333&$2.95^{+0.73}_{-2.95}$&736&$5.60^{+1.74}_{-5.60}$&344\\
3C 433&$0.96^{+0.76}_{-0.27}$&$0.240^{+0.117}_{-0.070}$&S&67&$8.18^{+2.52}_{-5.65}$&34&$30.2^{+2.1}_{-7.4}$&31\\
3C 452&$1.32^{+0.10}_{-0.08}$&$0.788^{+0.052}_{-0.070}$&E&211&$6.84^{+0.98}_{-1.31}$&105&$28.3^{+2.1}_{-1.8}$&94\\
3C 452&&&W&221&$6.69^{+0.99}_{-1.30}$&110&$26.1^{+1.9}_{-1.7}$&95\\
PKS 0034$-$01&0.61&0.0632&C&42&$<$8.30&21&$<$11.8&24\\
PKS 0038+09&$1.82^{+0.12}_{-0.18}$&$2.65^{+0.60}_{-0.78}$&C&114&$56.2^{+18.4}_{-11.9}$&57&$136^{+38}_{-28}$&71\\
PKS 0043$-$42&$1.59^{+0.96}_{-0.33}$&$1.32^{+0.28}_{-0.40}$&N&158&$19.6^{+1.8}_{-1.8}$&86&$36.6^{+2.7}_{-2.6}$&34\\
PKS 0043$-$42&&&S&162&$19.0^{+1.8}_{-1.8}$&105&$29.7^{+2.5}_{-2.1}$&28\\
PKS 0213$-$13&$0.85^{+0.09}_{-0.15}$&$0.205^{+0.080}_{-0.101}$&C&116&$5.44^{+1.99}_{-3.70}$&58&$20.6^{+6.9}_{-5.7}$&69\\
PKS 0349$-$27&$0.86^{+0.16}_{-0.20}$&$0.332^{+0.176}_{-0.246}$&S&294&$3.64^{+1.55}_{-1.71}$&159&$6.28^{+1.60}_{-0.92}$&79\\
PKS 0404+03&0.93&0.268&N&280&$<$2.90&149&$<$6.84&76\\
PKS 0404+03&&&S&298&$<$2.66&178&$<$5.41&73\\
PKS 0806$-$10&0.86&0.206&C&131&$<$6.53&65&$<$14.7&49\\
PKS 0945+07&$1.64^{+0.06}_{-0.10}$&$1.79^{+0.21}_{-0.31}$&W&186&$13.6^{+3.0}_{-2.1}$&107&$27.6^{+3.9}_{-3.5}$&64\\
PKS 1559+02&$0.65^{+0.10}_{-0.18}$&$0.790^{+0.160}_{-0.180}$&E&213&$4.12^{+0.99}_{-0.71}$&106&$8.25^{+1.48}_{-1.16}$&84\\
PKS 2221$-$02&$1.09^{+0.60}_{-0.40}$&$1.43^{+0.13}_{-0.22}$&S&325&$3.98^{+0.97}_{-0.52}$&162&$9.17^{+1.01}_{-0.76}$&98\\
PKS 2356$-$61&$1.19^{+0.14}_{-0.20}$&$1.38^{+0.25}_{-0.37}$&N&293&$8.75^{+1.33}_{-1.35}$&146&$22.5^{+1.9}_{-1.7}$&78\\
PKS 2356$-$61&&&S&308&$8.12^{+1.34}_{-1.37}$&154&$21.2^{+1.8}_{-1.7}$&78\\
\hline
3C  16&2.10&4.81&C&218&$<$54.6&109&$<$145&89\\
3C  46&$2.11^{+0.14}_{-0.20}$&$6.03^{+1.37}_{-1.72}$&C&586&$13.2^{+3.9}_{-6.3}$&293&$53.5^{+11.5}_{-6.9}$&167\\
3C  200&$1.74^{+0.16}_{-0.23}$&$2.64^{+0.93}_{-0.99}$&C&89&$103^{+32}_{-20}$&50&$261^{+56}_{-36}$&44\\
3C  228&$2.22^{+2.39}_{-0.71}$&$3.22^{+0.67}_{-0.89}$&C&170&$40.6^{+11.5}_{-10.8}$&98&$129^{+22}_{-15}$&63\\
3C  244.1&$2.05^{+0.17}_{-0.19}$&$4.54^{+1.36}_{-1.33}$&C&166&$94.6^{+15.9}_{-12.3}$&96&$189^{+29}_{-18}$&44\\
3C  274.1&$0.95^{+0.29}_{-0.23}$&$3.65^{+0.35}_{-0.29}$&C&485&$1.92^{+0.31}_{-0.71}$&274&$9.13^{+1.04}_{-0.95}$&197\\
3C  330&$1.61^{+1.26}_{-0.35}$&$4.64^{+0.70}_{-0.86}$&C&215&$42.2^{+6.4}_{-5.7}$&125&$101^{+9}_{-8}$&48\\
3C  427.1&$3.14^{+5.27}_{-1.18}$&$26.2^{+2.5}_{-2.5}$&C&104&$428^{+18}_{-21}$&60&$784^{+40}_{-45}$&44\\
3C  457&$3.10^{+2.95}_{-1.00}$&$6.14^{+1.05}_{-1.13}$&C&633&$9.49^{+1.15}_{-1.17}$&366&$18.0^{+1.7}_{-1.5}$&232\\
6C  0850+3747&$2.86^{+2.11}_{-0.78}$&$4.52^{+0.70}_{-0.77}$&S&163&$88.2^{+7.3}_{-8.7}$&95&$185^{+14}_{-12}$&59\\
6C  0857+3945&$1.69^{+0.24}_{-0.55}$&$2.87^{+1.62}_{-2.06}$&C&633&$6.54^{+5.15}_{-2.67}$&387&$13.0^{+8.6}_{-3.8}$&157\\
6C  1132+3439&$1.71^{+0.69}_{-0.37}$&$7.30^{+1.61}_{-1.84}$&C&299&$41.0^{+4.7}_{-3.6}$&176&$69.3^{+6.4}_{-5.7}$&90\\
6C  1200+3416&$2.46^{+1.89}_{-0.65}$&$8.28^{+0.95}_{-0.92}$&C&171&$145^{+9}_{-8}$&110&$173^{+13}_{-17}$&58\\
7C  0219+3423&$1.36^{+0.18}_{-0.91}$&$1.27^{+0.71}_{-1.25}$&C&111&$50.6^{+35.1}_{-25.8}$&62&$117^{+62}_{-36}$&75\\

\hline\end{tabular}\begin{flushleft}

$^{a}$ Bolometric X-ray luminosity within $R_{500}$.
$^{b}$ North, South, East, West and Combined lobes.
$^{c}$ The quoted volume is the radius of a sphere of the same volume as the lobe.
\end{flushleft}\end{minipage}\end{footnotesize}\label{tab:lobes}\end{table*}

%% file: tab_flux.tex
\begin{table*}\centering
\caption{FRII lobe flux densities.}
\begin{footnotesize}
\begin{tabular}{llcccccc}\hline\hline
Source&Lobe&Radio&Radio&Method$^{a}$&Photon&1~keV~X-ray&$\chi^2$/dof\\
&&Frequency&Flux density&&Index&flux density&\\
&&(GHz)&(Jy)&&&(nJy)&\\
\hline
3C 33&N&1.48&1.44&S&$1.14^{+0.71}_{-0.54}$&$3.98^{+0.78}_{-0.78}$&2.4/5\\
3C 33&S&1.48&1.61&S&$1.45^{+0.94}_{-0.77}$&$4.10^{+0.99}_{-0.99}$&6.1/4\\
3C 35&N&1.42&0.465&S&$1.77^{+0.43}_{-0.37}$&$27.4^{+5.6}_{-5.2}$&16/13\\
3C 35&S&1.42&0.532&C&$1.43^{+0.24}_{-0.23}$&$11.6^{+2.6}_{-2.6}$&1.8/6\\
3C 98&N&0.178&29.2&S&$1.97^{+0.46}_{-0.40}$&$12.6^{+2.5}_{-2.3}$&11/7\\
3C 192&C&1.41&4.09&S&$1.26^{+0.52}_{-0.45}$&$15.8^{+3.6}_{-3.4}$&4.9/5\\
3C 219&N&1.52&2.85&S&$1.64^{+0.14}_{-0.14}$&$17.1^{+1.3}_{-1.3}$&15/14\\
3C 219&S&1.52&2.99&S&$1.77^{+0.20}_{-0.19}$&$11.4^{+1.0}_{-1.0}$&7.5/8\\
4C 73.08&W&0.609&3.95&S&$1.44^{+0.45}_{-0.40}$&$34.1^{+4.6}_{-4.6}$&19/23\\
3C 285&E&1.65&0.790&S&$1.53^{+0.62}_{-0.63}$&$2.53^{+0.51}_{-0.53}$&1.4/2\\
3C 285&W&1.65&0.857&S&$1.79^{+0.67}_{-0.72}$&$3.28^{+0.57}_{-0.56}$&0.2/2\\
3C 303&C&1.45&0.900&F&1.50&$0.891^{+0.425}_{-0.425}$&\\
3C 321&N&1.51&0.135&M&1.50&$0.359^{+0.207}_{-0.207}$&\\
3C 321&S&1.51&1.32&M&1.50&$1.28^{+0.27}_{-0.27}$&\\
3C 326&E&1.40&1.87&S&$1.41^{+0.68}_{-0.71}$&$27.2^{+6.9}_{-6.9}$&27/18\\
3C 326&W&1.40&1.44&F&1.50&$16.2^{+4.1}_{-4.1}$&\\
3C 433&S&0.178&53.9&S&$2.10^{+0.71}_{-0.54}$&$3.19^{+0.68}_{-0.62}$&1.7/3\\
3C 452&E&1.41&4.25&S&$1.47^{+0.15}_{-0.20}$&$13.2^{+0.9}_{-0.9}$&32/53\\
3C 452&W&1.41&3.97&S&$1.58^{+0.17}_{-0.26}$&$12.8^{+0.9}_{-0.9}$&61/52\\
PKS 0034$-$01&C&0.408&9.74&S&$1.94^{+0.22}_{-0.21}$&$5.00^{+0.50}_{-0.50}$&6.3/5\\
PKS 0038+09&C&0.408&11.5&F&1.50&$8.60^{+1.65}_{-1.66}$&\\
PKS 0043$-$42&N&0.408&8.25&U&1.50&$<$0.990&\\
PKS 0043$-$42&S&0.408&7.40&U&1.50&$<$0.788&\\
PKS 0213$-$13&C&0.408&11.7&M&1.50&$4.00^{+0.65}_{-0.65}$&\\
PKS 0349$-$27&S&1.47&1.03&M&1.50&$10.9^{+1.1}_{-1.1}$&\\
PKS 0404+03&N&0.178&11.1&U&1.50&$<$4.66&\\
PKS 0404+03&S&0.178&8.27&U&1.50&$<$4.15&\\
PKS 0806$-$10&C&0.408&10.2&U&1.50&$<$2.74&\\
PKS 0945+07&W&1.43&2.29&S&$1.85^{+0.20}_{-0.19}$&$11.5^{+0.9}_{-0.9}$&19/15\\
PKS 1559+02&E&0.408&10.8&S&$1.56^{+0.45}_{-0.39}$&$6.07^{+1.42}_{-1.35}$&3.4/6\\
PKS 2221$-$02&S&1.42&3.40&S&$1.31^{+0.33}_{-0.31}$&$11.6^{+1.9}_{-1.9}$&23/23\\
PKS 2356$-$61&N&1.47&3.81&S&$1.58^{+0.25}_{-0.23}$&$17.7^{+2.2}_{-2.2}$&11/14\\
PKS 2356$-$61&S&1.47&5.40&S&$1.44^{+0.25}_{-0.24}$&$19.0^{+2.4}_{-2.4}$&13/16\\
\hline
3C  16&C&1.54&1.54&M&1.50&$2.26^{+1.09}_{-1.09}$&\\
3C  46&C&1.48&0.732&M&1.50&$4.99^{+1.62}_{-1.63}$&\\
3C  200&C&1.49&1.30&M&1.50&$1.76^{+0.69}_{-0.69}$&\\
3C  228&C&1.44&2.54&M&1.50&$3.24^{+0.59}_{-0.59}$&\\
3C  244.1&C&1.44&2.31&M&1.50&$1.52^{+1.01}_{-1.01}$&\\
3C  274.1&C&1.44&1.10&S&$1.39^{+0.16}_{-0.16}$&$5.38^{+0.63}_{-0.63}$&8.7/11\\
3C  330&C&1.49&2.39&M&1.50&$1.10^{+0.24}_{-0.24}$&\\
3C  427.1&C&1.53&2.89&M&1.50&$1.85^{+0.34}_{-0.33}$&\\
3C  457&C&1.45&0.650&S&$1.59^{+0.18}_{-0.18}$&$6.08^{+0.71}_{-0.71}$&19/19\\
6C  0850+3747&S&1.41&0.181&M&1.50&$1.22^{+0.33}_{-0.34}$&\\
6C  0857+3945&C&1.41&0.266&M&1.50&$2.13^{+0.51}_{-0.51}$&\\
6C  1132+3439&C&1.44&0.390&M&1.50&$2.15^{+0.42}_{-0.42}$&\\
6C  1200+3416&C&1.41&0.154&M&1.50&$1.03^{+0.22}_{-0.22}$&\\
7C  0219+3423&C&1.41&0.0847&U&1.50&$<$0.924&\\
\hline\end{tabular}\begin{flushleft}

$^{a}$ S = fitted index, C = fitted index from combined lobe spectra, F = fit with fixed index, M = modelled with unbinned data, U = upper limit
\end{flushleft}\end{footnotesize}\label{tab:flux}\end{table*}

%% file: tab_BandP.tex
\begin{table*}\centering
\caption{FRII lobe equipartition and Inverse Compton fields and pressures}
\begin{footnotesize}
\begin{tabular}{llcccccc}\hline\hline
Source&Lobe$^a$&$B_{\rm eqp}$&$B_{\rm obs}$&$P_{\rm eqp}$&$P_{\rm obs}$&Mach no.&$Q_{\rm jet}$\\
&&$\times{10}^{-10}$~T&$\times{10}^{-10}$~T&$\times{10}^{-14}$~Pa&$\times{10}^{-14}$~Pa&&$\times{10}^{44}$~erg~s$^{-1}$\\
\hline
3C 33&N&5.61&$2.29^{+0.31}_{-0.23}$&8.34&$19.8^{+3.6}_{-3.5}$&$4.1\pm0.8$&$3.30\pm1.43$\\
3C 33&S&5.88&$2.40^{+0.42}_{-0.29}$&9.17&$21.8^{+4.9}_{-4.8}$&$3.5\pm0.8$\\
3C 35&N&1.75&$0.377^{+0.049}_{-0.039}$&0.814&$5.56^{+1.12}_{-1.04}$&$1.8\pm0.4$&$2.45\pm1.23$\\
3C 35&S&1.82&$0.679^{+0.112}_{-0.078}$&0.879&$2.42^{+0.53}_{-0.52}$&$1.3\pm0.4$\\
3C 98&N&7.87&$2.69^{+0.34}_{-0.27}$&16.4&$51.8^{+9.8}_{-9.0}$&$4.1\pm1.3$&$7.32\pm2.66$\\
3C 192&C&6.97&$1.85^{+0.29}_{-0.21}$&12.9&$61.8^{+13.7}_{-13.2}$&$3.3\pm0.5$&$4.69\pm2.52$\\
3C 219&N&7.48&$1.84^{+0.09}_{-0.08}$&14.8&$80.6^{+5.9}_{-6.0}$&$1.7\pm0.1$&$17.6\pm2.3$\\
3C 219&S&7.91&$2.40^{+0.14}_{-0.12}$&16.6&$63.6^{+5.5}_{-5.5}$&$1.5\pm0.1$\\
4C 73.08&W&2.03&$0.810^{+0.072}_{-0.058}$&1.09&$2.67^{+0.34}_{-0.33}$&$1.9\pm0.5$&$3.60\pm1.07$\\
3C 285&E&5.09&$2.28^{+0.34}_{-0.23}$&6.87&$14.1^{+2.6}_{-2.6}$&$1.1\pm0.1$&$1.27\pm0.35$\\
3C 285&W&4.44&$2.06^{+0.24}_{-0.18}$&5.24&$10.3^{+1.6}_{-1.5}$&$1.1\pm0.1$\\
3C 303&C&10.7&$4.92^{+2.29}_{-1.01}$&30.5&$60.5^{+26.3}_{-23.7}$&$1.5\pm0.3$&$1.32\pm1.22$\\
3C 321&N&10.7&$2.55^{+1.67}_{-0.60}$&30.6&$177^{+101}_{-99}$&$21.5\pm6.7$&$2.68\pm1.43$\\
3C 321&S&10.4&$4.65^{+0.68}_{-0.49}$&28.5&$58.5^{+11.0}_{-10.6}$&$12.8\pm6.7$\\
3C 326&E&2.02&$0.894^{+0.168}_{-0.111}$&1.08&$2.15^{+0.62}_{-0.39}$&$0.7\pm0.2$&$1.37\pm0.80$\\
3C 326&W&1.69&$1.04^{+0.19}_{-0.13}$&0.761&$1.01^{+0.19}_{-0.16}$&$0.7\pm0.2$\\
3C 433&S&22.7&$11.7^{+1.6}_{-1.3}$&137&$230^{+41}_{-36}$&$4.8\pm1.2$&$14.2\pm4.2$\\
3C 452&E&6.13&$2.20^{+0.09}_{-0.08}$&9.97&$29.2^{+1.8}_{-1.8}$&$1.9\pm0.2$&$6.45\pm1.00$\\
3C 452&W&5.97&$2.15^{+0.09}_{-0.08}$&9.46&$27.5^{+1.8}_{-1.7}$&$1.9\pm0.2$\\
PKS 0034$-$01&C&17.1&$3.89^{+0.25}_{-0.21}$&77.5&$481^{+47}_{-47}$&\\
PKS 0038+09&C&12.9&$3.85^{+0.52}_{-0.38}$&43.9&$173^{+32}_{-32}$&$1.6\pm0.3$&$57.2\pm28.4$\\
PKS 0043$-$42&N&16.1&$>$9.99&68.7&$<$90.4&$<$2.0&$<$2.95\\
PKS 0043$-$42&S&18.3&$>$10.9&88.7&$<$123&$<$2.3&\\
PKS 0213$-$13&C&11.5&$5.60^{+0.62}_{-0.48}$&34.9&$63.2^{+9.0}_{-8.6}$&$3.1\pm0.8$&$19.4\pm11.4$\\
PKS 0349$-$27&S&4.31&$1.05^{+0.07}_{-0.06}$&4.92&$27.2^{+2.6}_{-2.6}$&$2.5\pm0.6$&$3.58\pm0.89$\\
PKS 0404+03&N&7.47&$>$3.09&14.8&$<$34.5&\\
PKS 0404+03&S&6.81&$>$3.18&12.3&$<$23.8&\\
PKS 0806$-$10&C&12.3&$>$6.01&40.3&$<$72.9&\\
PKS 0945+07&W&7.33&$1.68^{+0.08}_{-0.07}$&14.3&$87.2^{+6.5}_{-6.7}$&$2.3\pm0.2$&$16.2\pm2.0$\\
PKS 1559+02&E&7.85&$3.78^{+0.60}_{-0.44}$&16.4&$30.2^{+6.3}_{-5.6}$&$2.5\pm0.3$&$8.27\pm2.01$\\
PKS 2221$-$02&S&4.52&$1.98^{+0.22}_{-0.17}$&5.42&$11.6^{+1.8}_{-1.7}$&$1.6\pm0.2$&$1.60\pm0.30$\\
PKS 2356$-$61&N&7.71&$1.82^{+0.15}_{-0.12}$&15.8&$92.2^{+11.4}_{-11.3}$&$2.9\pm0.3$&$28.6\pm6.4$\\
PKS 2356$-$61&S&8.47&$2.15^{+0.17}_{-0.14}$&19.1&$98.6^{+12.2}_{-12.0}$&$3.1\pm0.3$\\
\hline
3C  16&C&12.6&$6.26^{+2.97}_{-1.30}$&42.2&$74.7^{+31.8}_{-38.8}$&\\
3C  46&C&6.52&$2.61^{+0.68}_{-0.40}$&11.3&$27.6^{+8.5}_{-8.2}$&$1.4\pm0.3$&$31.8\pm24.5$\\
3C  200&C&22.9&$7.59^{+2.60}_{-1.35}$&139&$463^{+178}_{-173}$&$2.0\pm0.4$&$91.1\pm80.5$\\
3C  228&C&23.1&$8.95^{+1.13}_{-0.85}$&141&$364^{+63}_{-61}$&$2.7\pm0.4$&$209\pm98$\\
3C  244.1&C&25.8&$11.5^{+10.6}_{-3.0}$&177&$366^{+226}_{-185}$&$1.8\pm0.5$&$34.9\pm44.1$\\
3C  274.1&C&6.21&$3.07^{+0.23}_{-0.19}$&10.2&$18.2^{+1.8}_{-1.8}$&$2.8\pm0.4$&$54.3\pm17.9$\\
3C  330&C&28.4&$17.5^{+2.7}_{-1.9}$&214&$284^{+45}_{-39}$&$2.4\pm0.2$&$42.2\pm15.3$\\
3C  427.1&C&33.0&$16.1^{+2.0}_{-1.5}$&290&$526^{+83}_{-80}$&$1.1\pm0.1$&$84.4\pm29.7$\\
3C  457&C&4.76&$2.12^{+0.13}_{-0.16}$&6.02&$12.5^{+1.3}_{-1.3}$&$1.1\pm0.1$&$34.4\pm9.0$\\
6C  0850+3747&S&9.78&$2.48^{+0.52}_{-0.33}$&25.4&$132^{+35}_{-35}$&$1.2\pm0.2$&$22.9\pm14.0$\\
6C  0857+3945&C&5.82&$2.66^{+0.46}_{-0.31}$&8.97&$18.0^{+3.9}_{-3.7}$&$1.5\pm0.5$&$20.0\pm14.8$\\
6C  1132+3439&C&9.32&$3.28^{+0.44}_{-0.33}$&23.1&$69.5^{+13.0}_{-12.8}$&$1.2\pm0.1$&$24.2\pm10.3$\\
6C  1200+3416&C&11.2&$2.98^{+0.45}_{-0.32}$&33.3&$159^{+33}_{-33}$&$1.0\pm0.1$&$27.5\pm13.0$\\
7C  0219+3423&C&8.32&$>$2.43&18.4&$<$74.9&$<$1.2&$<$42.0\\

\hline\end{tabular}\begin{flushleft}

Column 1: source name. Col. 2: lobe. Col. 3: equipartition magnetic field. Col. 4: observed (inverse Compton) magnetic field. Col. 5: equipartition lobe pressure. Col. 6: observed (inverse Compton) lobe pressure. Col. 7: Mach number. Col. 8: Jet power (summed across lobes)

$^a$ North, South, East, West and Combined lobes. 
\end{flushleft}\end{footnotesize}\label{tab:BandP}\end{table*}

%% file: tab_KTau.tex
\begin{table*}
\centering
\caption{Correlation analysis using Kendall's $\tau$ tests generalised to include censored data, and partial generalised Kendall's $\tau$ tests taking account of a common correlation with redshift \citep{aks96}.}

\begin{footnotesize}
\begin{tabular}{p{5cm}llcc}\hline\hline
 &Sub-sample&$N$&$\tau/\sigma$&$p$\\
\hline
\multicolumn{5}{l}{Mach no. vs Radio luminosity, no redshift correlation}\\
&All&43&-0.69&0.49\\
&HERG&34&-1.10&0.27\\
&no 3C~321&32&-0.66&0.51\\
\multicolumn{5}{l}{Mach no. vs ICM luminosity, no redshift correlation}\\
&All&43&-3.75&0.0002\\
&HERG&34&-3.32&0.001\\
&no 3C~321&32&-2.84&0.005\\
\multicolumn{5}{l}{ICM luminosity vs lobe-tip external pressure, with a redshift correlation}\\
&All&43&3.20&0.001\\
&HERG&34&3.48&0.0005\\
&LERG&9&3.20&0.001\\
\multicolumn{5}{l}{Mach no. vs ICM luminosity, (no estimated temperatures), no redshift correlation}\\
&All&24&2.20&0.028\\
\multicolumn{5}{l}{Jet power vs radio luminosity, with a redshift correlation}\\
&All&33&4.11&$<0.0001$\\
\multicolumn{5}{l}{Lobe volume vs length, no redshift correlation}\\
&All&37&5.30&$<0.0001$\\
\hline\end{tabular}

$N$ is sample size; $\tau$ is the correlation statistic; $\sigma$ is the standard deviation; $p$ is probability under the null hypothesis.
\end{footnotesize}\label{tab:KTau}\end{table*}